\documentclass[prb,twocolumn,notitlepage,superscriptaddress]{revtex4-2}
\usepackage{CJK}
\usepackage{amsfonts}
\usepackage{amsmath}
\usepackage{ dsfont }
\usepackage{hyperref}
\usepackage{amssymb}
\usepackage{xcolor}
\usepackage{xspace}
\usepackage{ragged2e}
\usepackage{relsize}%\mathlarger command
\usepackage[bottom]{footmisc}

\usepackage{tikz}
\usetikzlibrary{calc}
\newcommand{\nocontentsline}[3]{}
\newcommand{\tocless}[2]{\bgroup\let\addcontentsline=\nocontentsline#1{#2}\egroup}
\newcommand{\be}{\begin{equation}}
\newcommand{\ee}{\end{equation}}
\newcommand{\bea}{\begin{equation} \begin{aligned}}
\newcommand{\eea}{\end{aligned} \end{equation} }
\newcommand{\bi}{\begin{itemize}}
\newcommand{\ei}{\end{itemize}}

\newcommand{\la}{\lambda}
\renewcommand{\be}{\beta}
\newcommand{\al}{\alpha}
\newcommand{\bpm}{\begin{pmatrix}}
\newcommand{\epm}{\end{pmatrix}}
\newcommand{\eps}{\epsilon}

\renewcommand{\th}{\theta}

\newcommand{\lp}{\left(}
\newcommand{\rp}{\right)}

\newcommand{\del}{\partial}

\newcommand{\Tr}{\text{Tr} \ }

\newcommand{\mbf}[1]{\mathbf{#1}}
\renewcommand{\d}{\downarrow}

\usepackage{color}
\usepackage{graphicx}
\usepackage{verbatim}
\usepackage{amsmath}
\usepackage{amssymb}
\usepackage{wasysym}
\usepackage{mathrsfs}
\usepackage[caption=false]{subfig}
\usepackage{url}
\usepackage{bbold}
\usepackage{slashed}
\usepackage{epstopdf}
\usepackage{braket}
\usepackage{float}
\usepackage[percent]{overpic}

\DeclareRobustCommand{\Sec}[1]{Sec.~\ref{#1}}

\DeclareRobustCommand{\App}[1]{App.~\ref{#1}}

\DeclareRobustCommand{\Tab}[1]{Table~\ref{#1}}

\DeclareRobustCommand{\Fig}[1]{Fig.~\ref{#1}}

\DeclareRobustCommand{\Eq}[1]{Eq.~(\ref{#1})}

\DeclareMathAlphabet\mathbfcal{OMS}{cmsy}{b}{n}

\RequirePackage[normalem]{ulem} %DIF PREAMBLE
\RequirePackage{color}\definecolor{RED}{rgb}{1,0,0}\definecolor{BLUE}{rgb}{0,0,1} %DIF PREAMBLE
\begin{document}

\title{Heavy Fermions as an Efficient  Representation of Atomistic Strain and Relaxation in Twisted Bilayer Graphene}

\author{Jonah Herzog-Arbeitman}
\thanks{These authors contributed equally.}
\affiliation{Department of Physics, Princeton University, Princeton, New Jersey 08544, USA}

\author{Jiabin Yu}
\thanks{These authors contributed equally.}
\affiliation{Department of Physics, Princeton University, Princeton, New Jersey 08544, USA}

\author{Dumitru C\u{a}lug\u{a}ru}
\affiliation{Department of Physics, Princeton University, Princeton, New Jersey 08544, USA}

\author{Haoyu Hu}
\affiliation{Donostia International Physics Center, P. Manuel de Lardizabal 4, 20018 Donostia-San
Sebastian, Spain}

\author{Nicolas Regnault}
\affiliation{Department of Physics, Princeton University, Princeton, New Jersey 08544, USA}
\affiliation{Laboratoire de Physique de l’Ecole normale sup\'erieure,
ENS, Universit\'e PSL, CNRS, Sorbonne Universit\'e,
Universit\'e Paris-Diderot, Sorbonne Paris Cit\'e, 75005 Paris, France}

\author{Oskar Vafek}
\affiliation{National High Magnetic Field Laboratory, Tallahassee, Florida, 32310, USA}
\affiliation{Department of Physics, Florida State University, Tallahassee, Florida 32306, USA}

\author{Jian Kang}
\affiliation{School of Physical Science and Technology, ShanghaiTech University, Shanghai 200031, China}

\author{B. Andrei Bernevig}
\email{bernevig@princeton.edu}
\affiliation{Department of Physics, Princeton University, Princeton, New Jersey 08544, USA}
\affiliation{Donostia International Physics Center, P. Manuel de Lardizabal 4, 20018 Donostia-San Sebastian, Spain}
\affiliation{IKERBASQUE, Basque Foundation for Science, Bilbao, Spain}

% \author{Jonah Herzog-Arbeitman$^*$, Jiabin Yu$^*$, Dumitru Calugaru, Haoyu Hu, Oskar Vafek, Nicolas Regnault, Jian Kang,B. Andrei Bernevig}

\date{\today}
\begin{abstract}
Although the strongly interacting flat bands in twisted bilayer graphene (TBG) have been approached using the minimal Bistritzer-MacDonald (BM) Hamiltonian, there is mounting evidence that strain and lattice relaxation are essential in correctly determining the order of the correlated insulator groundstates. These effects can be incorporated in an enhanced continuum model by introducing additional terms computed from the relaxation profile. To develop an analytical and physical understanding of these effects, we include strain and relaxation in the topological heavy fermion (HF) model of TBG. We find that strain and relaxation are very well captured in first order perturbation theory by projection onto the fully symmetric HF Hilbert space, and remarkably do not alter the interacting terms in the periodic Anderson model. Their effects are fully incorporated in the single-particle HF Hamiltonian, and can be reproduced in a minimal model with only 4 symmetry-breaking terms. Our results demonstrate that the heavy fermion framework of TBG is an efficient and robust representation of the perturbations encountered in experiment. 
\end{abstract}
\maketitle

\section{Introduction}

Flat bands have figured prominently in the theoretical study of magic angle twisted bilayer graphene ever since their prediction \cite{2010NanoL..10..804T,PhysRevLett.99.256802,PhysRevResearch.1.013001,2011PNAS..10812233B}. Most notably, the appearance of correlated insulators in their phase diagram can be explained from generalized models of flat band ferromagnetism, where exact Slater determinant groundstates and their excitations become computable in the perfectly flat, isolated limit \cite{PhysRevLett.122.246401,2021PhRvB.103t5414L,PhysRevB.103.205415,Bultinck2019GroundSA}. However, this strong coupling approach faces shortcomings in comparison to experiment \cite{PhysRevLett.129.117602,2022PhRvL.129n7001H,2023Natur.620..525N,2024arXiv240314078L}, where the ubiquitous effects of strain and lattice relaxation increase the dispersion of the flat bands and decrease their gap to remote bands \cite{PhysRevB.96.075311,PhysRevLett.127.027601,PhysRevB.100.035448,PhysRevX.11.041063,2022arXiv221102693W,PhysRevLett.128.156401,2024TDM....11c5015C}. The low energy manifold of states appearing in this limit is split by strain and relaxation, which in turn dictate which phase is selected at integer filling \cite{PhysRevX.11.041063,2023Natur.620..525N,2023Natur.623..942K}. A more robust framework is required to describe the low energy degrees of freedom in the presence of these unavoidable perturbations. 

Motivated a multitude of features in the strong coupling excitation spectra, Ref. \cite{2022PhRvL.129d7601S} derived a mapping between the low-energy moir\'e bands of the Bistritzer-MacDonald (BM) Hamiltonian and a topological heavy fermion model \cite{2011PNAS..10812233B}, featuring trivial, impurity-like $f$-modes with no kinetic energy coupled to two sets of Dirac fermions ($c$-modes). 
This mapping preserves all symmetries of the original Hamiltonian (including an emergent particle-hole symmetry \cite{2018arXiv180710676S} and global $U(4) \times U(4)$ limits \cite{2021PhRvB.103t5413B,Bultinck2019GroundSA}), and is a unitary change of basis within the \emph{low-energy} Hilbert space, including the flat bands as well as the low-lying states in the nearby dispersive bands. In this way, one obtains a generalized periodic Anderson model \cite{PhysRev.124.41,1966PhRv..149..491S}. This model numerically reproduces the large number of low-lying symmetry-broken correlated insulator groundstates that emerge in Hartree-Fock at integer fillings \cite{PhysRevLett.124.097601,PhysRevB.103.035427,PhysRevB.102.045107,PhysRevB.98.081102,PhysRevResearch.3.013033,PhysRevX.11.011014,PhysRevB.102.035161,PhysRevB.106.245129}. Moreover, one finds that the $f$-modes dominate the Hartree-Fock order parameter, yielding reliable analytical approximations for the groundstate and the excitation bands. An intuitive picture of the cascade physics emerges from these calculations, with heavy $f$-modes polarizing at integer fillings on top of the light $c$-mode Fermi sea \cite{PhysRevLett.127.266402}. This picture is borne out by effective field theory and dynamical mean-field theory calculations \cite{2023PhRvL.131b6502H,2023arXiv230104661Z,2023arXiv230908529R,2023NatCo..14.5036D,2024arXiv240214057C} and is compatible with experiment \cite{2021Natur.592..214R,2024arXiv240211749L,2024arXiv240212296B}. The analogy to Kondo physics \cite{PhysRevLett.131.026501,2023arXiv230302670L,2024PhRvB.109f4517W} has opened a new direction in the study of TBG and beyond \cite{2024arXiv240407253H,2024arXiv240211824S}, especially away from integer filling. 

However, the integer-filling groundstates of the projected BM model are not robust to strain, which disrupts the approximate $U(4) \times U(4)$ symmetry that organizes the strong coupling groundstates in the flat band limit \cite{PhysRevLett.122.246401,2021PhRvB.103t5414L,PhysRevB.103.205415,Bultinck2019GroundSA} and allows competing states near the low-energy manifold, such as the Incommensurate Kekule Spiral (IKS)\cite{PhysRevX.11.041063,PhysRevLett.128.156401,2023PhRvB.108w5128W}, to be selected. As we will show in this work, the expanded Hilbert space of the HF model easily accommodates strain and relaxation while preserving the heavy-fermion character of the bands. We make the case that heavy fermions provide a unifying formalism for TBG (and its multi-layer extensions \cite{2019PhRvB.100h5109K,2021arXiv211111060L,2023PhRvB.108c5129Y}) which survives the inevitable perturbations in experiments \cite{2021NatMa..20..956K,yoo2019atomic,2023Natur.620..525N,2023PNAS..12007151W,zhang2022correlated}. The development of a single-particle Heavy Fermion (HF) model\cite{2022PhRvL.129d7601S} including strain and relaxation is the focus of this paper. We study the interacting problem in coming work \cite{2025arXiv250208700H}.

Both strain and relaxation can be incorporated in the BM model. Early work \cite{PhysRevB.96.075311,PhysRevB.100.035448,PhysRevLett.127.126405,PhysRevLett.120.156405,PhysRevX.11.041063,PhysRevLett.128.156401,2022arXiv220908204W,2022APS..MARG56007W,PhysRevLett.127.027601,PhysRevX.11.041063} approached these effects with a minimalist approach coupling strain as a pseudo-gauge field to the Dirac fermions \cite{2010PhR...496..109V} and implementing relaxation through renormalized parameters and $\mbf{k}$-dependent (so-called ``non-local") interlayer tunneling. More recently, Kang and Vafek have shown that a generalized BM model can be derived systematically in a gradient expansion given a moir\'e relaxation profile \cite{PhysRevB.107.075408}, possibly including strain \cite{KV_toappear}. This formalism yields a moir\'e model that perfectly matches atomistic calculations, albeit requiring many parameters beyond the typical BM model --- as expected in a microscopic Hamiltonian. Here we show that the HF model can quantitatively reproduce the Kang-Vafek (KV) model with only a handful of parameters (four in a minimal model), yielding a simple yet microscopically accurate Hamiltonian amenable to theoretical study. 

This paper is organized as follows. In \Sec{sec:BM}, we lay out our conventions by reviewing the BM model and the original HF model near $\th = 1.05^\circ$ and discussing their symmetries. In \Sec{sec:Kaxiras}, we introduce corrections to the BM model from uniaxial heterostrain and non-local tunneling, which break the crystalline symmetries and emergent particle-hole symmetry, respectively, leaving only $C_{2z}\mathcal{T}$. \Sec{sec:hfminimal} contains the resulting corrections to the HF model, and shows that the these corrections can be incorporated as symmetry-breaking kinetic terms acting on an unaltered (fully symmetric) Hilbert space of $f$- and $c$-modes. This is analogous to the treatment of displacement field in twisted trilayer graphene \cite{2023PhRvB.108c5129Y}. Lastly, in \Sec{sec:hffullrelax} we obtain a heavy fermion representation for the  Kang-Vafek model \cite{PhysRevB.107.075408}, which is an extension of the BM model incorporating microscopically accurate lattice relaxation. This model provides an extremely accurate match to atomistic band structures \cite{2010NanoL..10..804T,PhysRevLett.99.256802,PhysRevResearch.1.013001} and shows significant deviations from the typical BM flat bands. We find that this complex band structure and its patterns of symmetry-breaking is nevertheless captured by a generalized HF model on a fully symmetric Hilbert space. We conclude in \Sec{sec:conclusion} with comments on the applicability of the HF model to the strongly correlated TBG problem. 

\section{Bistritzer-MacDonald and Heavy Fermion Model}
\label{sec:BM}

In this section, we review the BM model, the HF model obtained from it, and their symmetries. 

\subsection{Bistritzer-MacDonald Model}
\label{sec:BMmainintro}

We first set our conventions while briefly summarizing the BM model. We take monolayer graphene to have lattice vectors $\mbf{a}_{G,i}, |\mbf{a}_{G,i}| = a = 0.246$nm, yielding Dirac cones with velocity $v_F = 5.944\text{eV \AA}$ (here and throughout, we set $\hbar = 1$) at the graphene $K_G$ point 
\bea
\mbf{K}_G &= \frac{4\pi}{3 a} \hat{x}
\eea
and its $C_{2z}$-related partner $K'_G$. Throughout, $C_{nz}$ denotes a $n$-fold rotation about the $z$-axis. All $C_{3z}$-related $K_G$ points are equivalent in the graphene Brillouin zone (BZ). Twisting two layers of graphene by $\pm \th/2$ allows coupling between the low-energy states at nearby $K$ points $R(\pm \frac{\th}{2}) \mbf{K}_G$ (and similarly for $K'$) where $R(\th)$ is a rotation matrix. The momentum transfer between the $K$ points is
\bea
\mbf{q}_1 =  R(-\frac{\th}{2}) \mbf{K} - R(\frac{\th}{2}) \mbf{K} = -\frac{4\pi}{3a} \, 2 \sin \frac{\th}{2} \, \hat{y}
\eea
with its $C_3$-related partners $\mbf{q}_{j+1} = C_3 \mbf{q}_j$.  We then define the moir\'e reciprocal lattice vectors $\mbf{b}_j$ and real space lattice vectors $\mbf{a}_i$ by
\bea
\mbf{b}_j = \mbf{q}_3 - \mbf{q}_j, \quad \mbf{a}_i \cdot \mbf{b}_j = 2\pi \delta_{ij}
\eea
which define the moir\'e BZ and moir\'e unit cell respectively. 

At low-energy, scattering between the $K_G$ and $K'_G$ points (which are far away in momentum space) is suppressed and there is an emergent valley quantum number $\eta = \pm 1 = K_G,K'_G$ which labels the states. Since the $K_G$ and $K'_G$ points are related by spinless time-reversal, we can focus on the model in the $K_G$ valley, denoted $h^K_{BM}$. The resulting Bistritzer-MacDonald model consists of the Dirac cones in each layer coupled by inter-layer tunneling $T(\mbf{r})$:
\bea
\label{eq:Hbmorigmain}
h^K_{BM}(\mbf{r}) = \bpm - i v_F \pmb{\nabla} \cdot  \pmb{\sigma}   & T(\mbf{r}) \\ T^\dag(\mbf{r}) & - i v_F \pmb{\nabla} \cdot  \pmb{\sigma}  \epm  
\eea
where we have dropped $O(\th)$ particle-hole breaking kinetic terms, which are negligble in comparison to the strain and relaxation terms that we will add shortly. The form of the tunneling matrix derived is
\bea
\label{eq:Tmain}
T(\mbf{r}) &= \sum_j e^{i \mbf{q}_j \cdot \mbf{r}} T_j 
\eea
and the $2 \times 2$ matrices $T_j$ acting on the A/B sublattice degree of freedom are $T_{j+1} = w_0 \sigma_0 + w_1(\sigma_1 \cos \frac{2\pi}{3}j + \sigma_2 \sin \frac{2\pi}{3}j)$. Originally, a two-center approximation yielded $w_0 = w_1 \sim 110$meV \cite{2011PNAS..10812233B}, neglecting variation within the moir\'e unit cell as well as $\mbf{k}$-dependence in the inter-layer tunneling. An improvement of the model was obtained shortly thereafter by taking $w_0/w_1 \sim 0.8$ \cite{PhysRevX.8.031087}, which physically reflects the different tunneling amplitudes at the AA versus AB stacking region in the moir\'e unit cell. This is the simplest model of relaxation, now understood to be extremely important for accurate modeling of moir\'e materials. Restoring the $\mbf{k}$-dependence in the interlayer coupling breaks the particle-hole symmetry of spectrum, and leads to better agreement with tight-binding calculations in the full moir\'e unit cell at commensurate angles \cite{PhysRevResearch.1.013001,2021PhRvB.103t5412S}. Finally, \Eq{eq:Hbmorigmain} assumes an ideal device where the two graphene monolayers layers survive their stacking undeformed. It is now clear that heterostrain is ubiquitous in TBG devices and must be included to capture in the enlarged dispersion that disrupts the flat bands. We will revisit the physics of strain, realistic $\mbf{k}$-dependent coupling, and microscopic lattice relaxation with increasingly accurate models throughout this work. 

The second quantized BM model is
\bea
H^K_{BM} &= \int d^2r \, c^\dag_{l \al}(\mbf{r}) \, [h^K_{BM}(\mbf{r})]_{l \al,l' \be} \, c_{l' \be}(\mbf{r}) \\
&= \sum_{\mbf{k},n} E_n(\mbf{k}) \gamma^\dag_{\mbf{k},n} \gamma_{\mbf{k},n}
\eea
which is diagonalized in the band basis $\gamma_{\mbf{k},n}^\dag$ for $\mbf{k}$ in the moir\'e BZ. A discussion of the momentum-space representation of $H^K_{BM}$ may be found in \App{app:OGBM}.

We now discuss the symmetries of the BM model. Graphene possesses $C_{6z}, C_{2x}$ and $\mathcal{T}$ crystalline symmetries (as well as translations), in addition to $SU(2)$ spin. At the $K_G$ point, the little group consists of $C_{3z}, C_{2z}\mathcal{T}$ and $C_{2x}$. In TBG, these three symmetries (along with moir\'e translation symmetry) are local to the $K_G$ valley and are preserved as a symmetries of the BM model in \Eq{eq:Hbmorigmain}, forming the wallpaper group  $p6'2'2$ (\#177.151 in the BNS setting).  There is also a unitary, anti-commuting particle-hole symmetry $P$ inherited from monolayer graphene, which takes $\mbf{r} \to -\mbf{r}$ \cite{2018arXiv180710676S}. \App{app:OGBM} contains the explicit representations of these symmetries.  

The commuting symmetries together protect nontrivial fragile topology \cite{2018arXiv180710676S,PhysRevX.9.021013,PhysRevB.99.195455,PhysRevB.100.195135,PhysRevB.99.155415} in the flat bands, and $\mathcal{P} = C_{2z}\mathcal{T} P, \mathcal{P}^2 = -1$ is sufficient to protect stable topology equivalent to having
\bea
\delta = 2 \mod 4
\eea
Dirac points \cite{2021PhRvB.103t5412S}. In fact, any $\mathcal{P}$-symmetric set of bands in TBG is topological with $\delta \neq 0$, meaning no lattice model preserving all symmetries of TBG is possible for any number of bands. This anomaly necessitates a different kind of low-energy model than the tight-binding approach typically taken. 

\begin{figure}[h]
\centering
\includegraphics[width=.9\columnwidth]{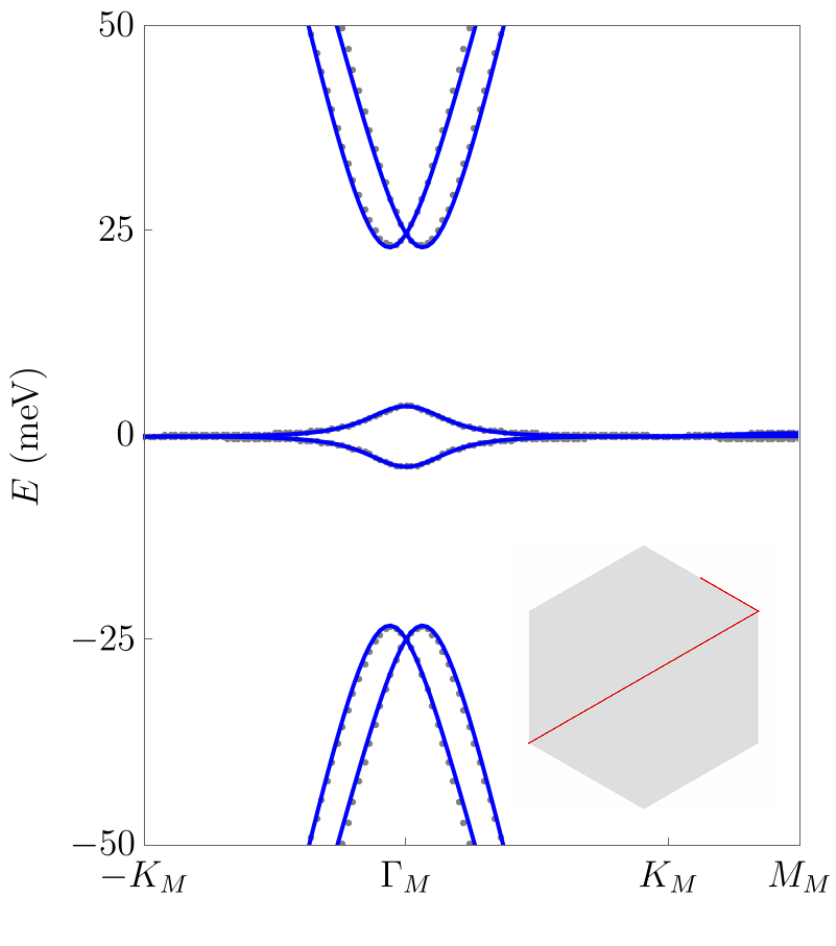}
\caption{Comparison of the BM model (gray dots) in \Eq{eq:Hbmorigmain} with the HF model (blue line)  in \Eq{eq:hHForiginalmain}, showing near-perfect agreement in the low-energy bands. We take $w_0/w_1 =0.8$ and $\th = 1.05^\circ$.
}
\label{fig:orig}
\end{figure}

\subsection{Heavy Fermion Model}

Since it is impossible to obtain fully symmetric, exponentially Wannier states (and hence a physical tight-binding model) for the two flat bands, we are motivated to include additional states in the Hilbert space. Ref. \cite{2022PhRvL.129d7601S} found that by mixing the low-lying states near $\Gamma$ in the dispersive bands with the flat band Hilbert space, it is possible to obtain a set of two $f$ modes (per valley per spin)
\bea
f^\dag_{\mbf{R},n} = \frac{1}{N} \sum_{\mbf{k},m =1,\dots,6} e^{- i \mbf{k} \cdot \mbf{R}} \gamma^\dag_{\mbf{k},m} \Omega^f_{mn}(\mbf{k})
\eea
where $\gamma^\dag_{\mbf{k},m}, m = 1,\dots,6$ are the 6 lowest energy BM model creation operators (the two flat bands and 4 dispersive bands shown in \Fig{fig:orig}),  $\Omega^f(\mbf{k})$ is a $6 \times 2$ matrix obeying $\Omega^{f \dag} \Omega^f = \mathbb{1}_{2\times 2}$ which disentangles the Wannier states $f^\dag_{\mbf{R},n}, n = 1,2$ from the set of low-energy bands, and $N$ is the number of $\mbf{k}$ points in the BZ. This disentanglement procedure can be performed numerically with Wannier90\cite{mostofi2014updated}, starting from gaussian $p_x-p_y$ trial wavefunctions, such that $f^\dag_{\mbf{R},n}$ are maximally localized\cite{RevModPhys.84.1419} within the $\mbf{R}$-th unit cell. We will refer to the $f$-modes as heavy fermions. 

The symmetry anomaly in TBG prevents the remaining 4 bands from being Wannierized while locally representing its symmetries. Thus, we are forced to introduce $\delta = 2$ flavors of Dirac electrons $c^\dag_{\mbf{k},\mbf{G},n}, n =1,\dots,4$ with an unbounded spectrum, i.e. infinite moir\'e reciprocal lattice vectors $\mbf{G}$, to carry the anomaly. However, since we are only interested in the low-energy physics dominated by the $\mbf{G}=0$ shell (the higher shells have large kinetic energy $> 100$meV), we will take
\bea
c^\dag_{\mbf{k},\mbf{G}=0,n} &= \sum_{m=1,\dots,6} \gamma^\dag_{\mbf{k},m} \Omega^c_{mn}(0), \quad n = 1,\dots,4 
\eea
so that $\Omega(\mbf{k}) = [\Omega^c(0) \, | \, \Omega^f(\mbf{k})]$ is a $6\times 6$ unitary matrix at $\mbf{k}=0$. Away from $\mbf{k}=0$, $\Omega(\mbf{k})$ will be only approximately unitary, but due to the large dispersion of the conduction bands, the low-energy physics is not sensitive to this approximation \cite{2022PhRvL.129d7601S}. 

In practice, $\Omega^c(0)$ is determined from the four-dimensional nullspace of $\Omega^f(\mbf{k}=0)$ (note that the $f$ and $c$ wavefunctions are orthogonal at $\mbf{k}=0$, so $\Omega^f(\mbf{k}=0)^\dag \Omega^c(0) = 0$). To fix the remaining freedom in the basis of the  nullspace, gauge fixing is used\cite{2022PhRvL.129d7601S} so that the $c$-electrons form specific representations of $\Gamma$ point symmetries $C_{3z}, C_{2z}\mathcal{T}, C_{2x}$. This point group has three irreps 
\begin{center}
\begin{tabular}{c | c c c }
 & $1$ & $C_{3z}$ & $C_{2x}$ \\
 \hline
$\Gamma_1$ & $1$ & $1$ & $1$ \\
$\Gamma_2$ & $1$ & $1$ & $-1$ \\
$\Gamma_3$ & $2$ & $-1$ & $0$ \\
\end{tabular}
\end{center}
and the 6 low energy bands carry the representations $\Gamma_1 \oplus \Gamma_2 \oplus 2 \Gamma_3$ (the two-dimensional $\Gamma_3$ irreps are furnished by the dispersive bands) \cite{2018arXiv180710676S}. The $f$-modes form an atomic limit incorporating the $\Gamma_3$ irrep, leaving the four $c$-modes to carry the remaining $\Gamma_1 \oplus \Gamma_2 \oplus \Gamma_3$ irreps. Since there is only one of each irrep carried by the $c$-electrons, the $c$-modes can be fixed (up to an irrelevant overall phase) by specifying the symmetry representations. Their representations may be found in \App{app:HFreview}.

Having fixed the wavefunctions of the $c$- and $f$-modes, it is straightforward to compute their effective Hamiltonian by acting on the basis with $h^K_{BM}$. The result is\cite{2022PhRvL.129d7601S}
\bea
\label{eq:hHForiginalmain}
h_{HF}^K(\mbf{k}) &= \bpm 
 0& v \mbf{k} \cdot (\sigma_0, i \sigma_3) &  \gamma \sigma_0 + v' \mbf{k} \cdot \pmb{\sigma} \\
v \mbf{k} \cdot (\sigma_0, -i \sigma_3)& M \sigma_1 & 0\\
 \gamma \sigma_0 + v' \mbf{k} \cdot \pmb{\sigma}  & 0 & 0 \\
\epm
\eea
where the upper $4\times 4$ block acts on the 4 $c$-modes (we neglect higher shells for compactness of notation), the lower $2\times 2$ block acts on the 2 $f$-modes, and $\mbf{k}$ is in the first BZ. The velocity of both Dirac cones is $v=-4.3$ eV \AA, and they are coupled by a mass parameter $M = 3.7$meV. The $f$-modes have zero kinetic energy, but are coupled to the Dirac nodes via $\gamma = -24.8$ eV and $v'=1.6$ eV \AA. \Fig{fig:orig} shows the excellent match between the HF and BM band structures. Note that \Fig{fig:orig} and all subsequent numerical calculations include a damping factor 
\bea
\gamma \sigma_0 + v' \mbf{k} \cdot \pmb{\sigma} \to e^{- \lambda^2 |\mbf{k}|^2/2}(\gamma \sigma_0 + v' \mbf{k} \cdot \pmb{\sigma})
\eea
in the $f-c$ coupling where $\la^2$ is the the Wannier spread corresponding to $\la \sim0.3 a_M$ where $a_M = |\mbf{a}_1|$ is the moir\'e lattice constant. This factor arises from the soft matrix element of the localized $f$-mode with the conduction electrons \cite{2022PhRvL.129d7601S}. 

\subsection{Outlook}

Symmetries play an important role in identifying and obtaining the heavy fermion representation. However, adding heterostrain and relaxation will break all symmetries except $C_{2z}\mathcal{T}$. Despite this reduction, $C_{2z}\mathcal{T}$ alone is enough to protect the fragile topology of the flat bands and obstruct a tight-binding model with local symmetries. This means that the necessity of a heavy fermion framework for describing the flat bands with a local model remains. 

More important than these single-particle considerations is the interacting problem. For instance, strong coupling \cite{PhysRevLett.122.246401,2021PhRvB.103t5414L,PhysRevB.103.205415,Bultinck2019GroundSA}, momentum space Monte Carlo \cite{PhysRevX.12.011061,2023arXiv230414064H,2023PhRvB.107x1105Z,2022PhRvB.106r4517Z,PhysRevB.109.125404}, and projected Hartree-Fock calculations \cite{PhysRevLett.124.097601,PhysRevX.11.041063,PhysRevB.103.035427,PhysRevB.102.045107,PhysRevB.98.081102,PhysRevResearch.3.013033,PhysRevX.11.011014,PhysRevB.102.035161,PhysRevB.106.245129} encounter no fundamental difficulties in the flat band Hilbert space, despite its topology and lack of local description. Instead, strain and relaxation hinder the application of these methods because the active bands become less flat and less well isolated. As the gap becomes smaller, interactions increasingly mix the flat and dispersive bands, causing flat band approximations to fail and requiring projected calculations to include a larger set of active bands and increasing computational expense. On the other hand, the larger Hilbert space accessible by interactions is well suited to the HF framework, which already includes nearby states in the dispersive bands. The locality of the model allows powerful numerical tools like dynamical mean field theory (DMFT) to be applied \cite{2023arXiv230908529R}. 

We now introduce strain and relaxation to the BM model and generalize the HF framework to include these corrections. 

\section{Heterostrain and Non-local Tunneling in the BM Model}
\label{sec:Kaxiras}

We now introduce heterostrain to the BM model \cite{PhysRevB.96.075311,PhysRevLett.127.027601,PhysRevB.100.035448,PhysRevX.11.041063,2022arXiv221102693W,PhysRevLett.128.156401}. First we consider the $l$-th graphene layer. Uniaxial strain is defined by the linear transformations $\mbf{r} \to (1+ \mathcal{E}_l)\mbf{r}$ where the matrix $\mathcal{E}_l$ acting on the coordinates $(x,y)$ can be decomposed into the form
\bea
\mathcal{E}_l = \bpm\eps^l_+ + \eps^l_-&\eps^l_{xy} \\ \eps^l_{xy} & \eps^l_+ - \eps^l_- \epm
\eea
neglecting the anti-symmetric (rotational) part. The constant $\eps_+$ is interpreted as isotropic expansion/contraction, and $\eps_-,\eps_{xy}$ are the area-preserving anisotropic/shear strains. Typical strains in moir\'e graphene are of the order $\eps \sim0.1\%$, and thus we can always work to linear order in  $\mathcal{E}$. 

Strain deforms the $K_G$ points in the top and bottom layers, but does not gap the Dirac points because it preserves $C_{2z}$ (we assume uniaxial, constant strain) and $\mathcal{T}$. The deformed $K_G$ points are given by
\bea
\mbf{\tilde{K}}^l & =  \big( R(l \th/2) - \mathcal{E}_l^{T} \big) \mbf{K}_G
\eea
where $\mathcal{E}_l$ is the strain in the $l$th layer, and thus the moir\'e $\mbf{q}_j$ vectors become 
\bea
\mbf{\tilde{q}}_j &= \mbf{\tilde{K}}^b - \mbf{\tilde{K}}^t
\eea
which depends only on the heterostrain tensor
\bea
\mathcal{E} = \mathcal{E}_t-\mathcal{E}_b
\eea
which is parameterized by $\eps_{\pm},\eps_{xy}$ without the $l$ index. Thus we see that heterostrain, where $\mathcal{E}_t = - \mathcal{E}_b$, couples directly to the moir\'e lattice, whereas homostrain $\mathcal{E}_t=+\mathcal{E}_b$ is absent. From here forward, we focus on heterostrain $\mathcal{E}$. Clearly $\mathcal{E}$ preserves $C_{2z}\mathcal{T}$ since it is uni-axial, but generically breaks $C_3$ and $C_{2x}$. 

Strain also couples direct to the graphene Dirac cone as a gauge field, giving the BM Hamiltonian \cite{2011PNAS..10812233B,PhysRevB.96.075311,PhysRevB.100.035448}
\bea
\label{eq:Hbmorigmainstrain}
\tilde{h}^K_{BM}(\mbf{r}) = \bpm - i v_F (\pmb{\nabla} - \mbf{A}^t) \cdot  \pmb{\sigma}   & \tilde{T}(\mbf{r}) \\ \tilde{T}^\dag(\mbf{r}) & - i v_F (\pmb{\nabla}-\mbf{A}^b) \cdot  \pmb{\sigma}  \epm  
\eea
where we have dropped $O(\th), O(\mathcal{E})$ particle-hole breaking kinetic terms, $\tilde{T}(\mbf{r})$ simply replaces \Eq{eq:Tmain} with the $\mbf{\tilde{q}}_i$, and the strain pseudo-gauge field \cite{vozmediano2010gauge,guinea2010energy} is, in our conventions (see \App{app:strain}),
\bea
\label{eq:strainfieldA}
\mbf{A}^l  = \frac{\be}{a/\sqrt{3}}  (\eps_-^l , -\eps_{xy}^l) \
\eea
where $\be \approx 3.14$ is the strain coupling constant and $a$ is the graphene lattice constant. Note that the $O(\mathcal{E})$ kinetic terms are suppressed relative to \Eq{eq:strainfieldA} by a factor of $a/a_M \sim \th \ll 1$. For heterostrain, $\mbf{A}^t =  - \mbf{A}^b$ is anti-symmetric in layer, and $\tilde{h}^K_{BM}(\mbf{r})$ thus preserves particle-hole $P$. 

\begin{figure*}[th]
\centering
\includegraphics[width=2\columnwidth]{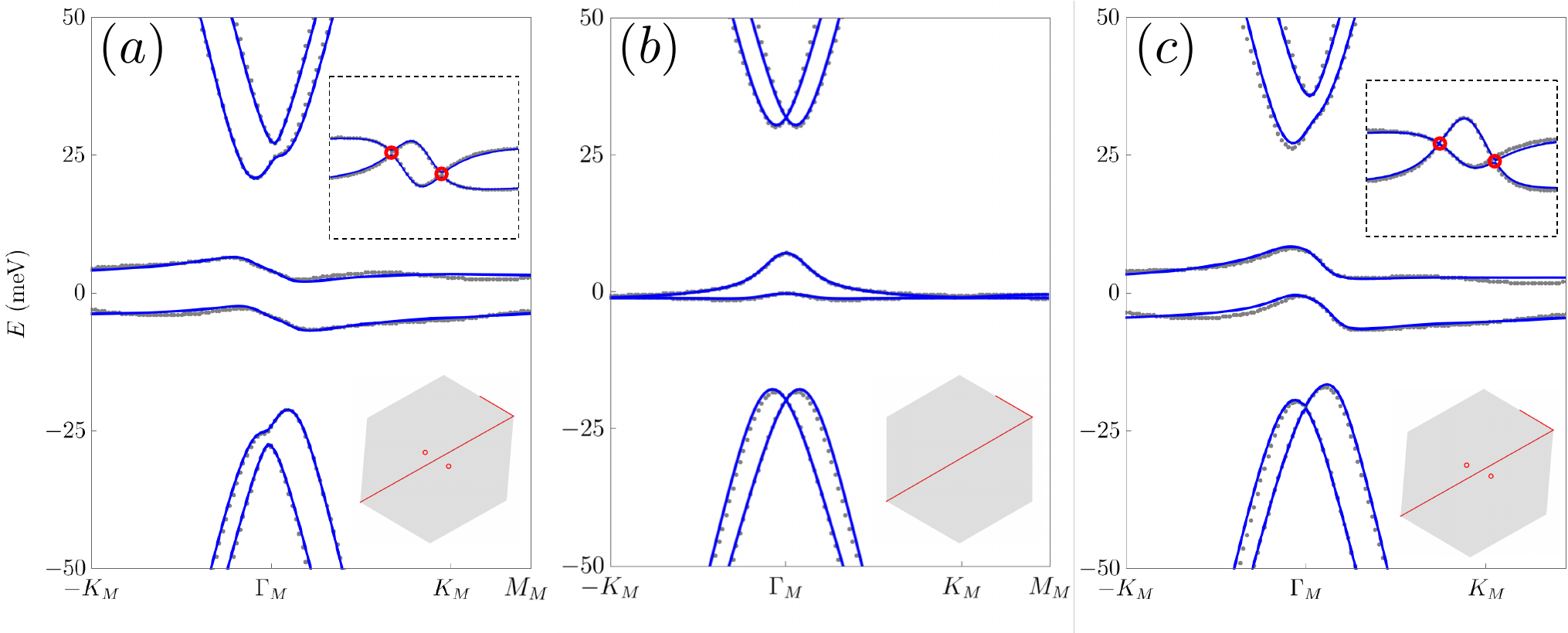}
\caption{Perturbations to the BM mode in \Sec{sec:hfminimal}. We consider $(a)$ heterostrain $\eps_{xy} = 0, \eps_\pm = \pm(\nu_G \mp 1) \eps/2$ with $\nu_G =0.16$ as the Poisson ratio and $\eps =0.002$ (inset: $C_{2z}\mathcal{T}$-protected nodes, marked by red circles in the BZ), $(b)$ non-local tunneling with $\la/a = 90$meV, and $(c)$ both heterostrain and non-local tunneling. The perturbed BM bands are shown in gray and the perturbed HF bands are shown in blue (including only the $\mbf{G}=0$ shell).
} 
\label{BM_3_fig}
\end{figure*}

The energy scale of the external symmetry breaking for $\eps \sim 0.001$ is $\sim 10$meV. This is a large energy compared to the bandwidth at magic angle $\th \sim 1.05^\circ$, but it is small compared to the energy range in which the heavy fermion mapping is valid, which is at least $\pm50$meV. As we will see in the following section, strain has a strong effect on the band structure (see \Fig{BM_3_fig}a), but does not disrupt the heavy fermion Hilbert space. 

The second effect we consider is the gradient inter-layer term\cite{PhysRevResearch.1.013001,PhysRevLett.127.196401,2019arXiv190300340W}, which is simply the inclusion of $\mbf{k}$-dependence in the interlayer coupling (see Ref. \cite{2021PhRvB.103t5412S} for a derivation in momentum space). Such a term takes the form
\bea
\label{eq:gradcoupling}
h_{grad} &= \bpm 0 & h.c. \\
\frac{1}{2}\{- i \pmb{\nabla}, \pmb{\Lambda}_{grad}\} & 0 \epm
\eea
where $\{\mbf{A},\mbf{B}\} = \mbf{A} \cdot \mbf{B}+\mbf{B} \cdot \mbf{A}$, and the moir\'e periodic coupling is
\bea
\pmb{\Lambda}_{grad}(\mbf{r}) &= - \la e^{i \mbf{q}_1 \cdot \mbf{r}} (\hat{x} \sigma_0+\hat{x} \sigma_1 + \hat{y} \sigma_2) + \dots \\
\eea
where the dots indicate $C_3$-related harmonics $\mbf{q}_2, \mbf{q}_3$, and $\la/a  = 90$meV is taken by Ref \cite{PhysRevResearch.1.013001}. We can see that $h_{grad}$ breaks $P$ because of the $-i \pmb{\nabla}$ factor (absent in the contact term $T(\mbf{r}$), which is odd under $\mbf{r} \to - \mbf{r}$. Otherwise, $h_{grad}$ preserves all crystalline symmetries. 

The order of magnitude of $h_{grad}$ is $\mbf{q}_1 \la \sim 10$meV. Much like the effects of strain, this perturbation has a strong effect on the dispersion of the flat bands (see \Fig{BM_3_fig}b), but is well within the $\pm 50$meV energy scale at which the HF approximation is valid. We will now show that both effects can be built into the HF model using simple perturbation theory. 

\section{Strain and Relaxation Corrections to the Heavy Fermion Model}
\label{sec:hfminimal}

Order of magnitude estimates indicate that the relevant perturbations to the BM model can be captured by perturbation theory within the unperturbed HF Hilbert space. This assertion implies that the perturbed HF model takes the form
\bea
\label{eq:Hfull}
h^K_{HF,\eps,\Lambda}(\mbf{k}) &= h^K_{HF}(\mbf{k}) + \delta h^K_\eps(\mbf{k}) + \delta h^K_\Lambda(\mbf{k}) 
\eea
acting on the \emph{unperturbed} $f$- and $c$-mode wavefunctions. Here the correction terms $ \delta h_\eps(\mbf{k}), \delta h_\Lambda(\mbf{k})$ correspond to strain and non-local tunneling. They are simply the first-order perturbation theory expressions
\bea
 \null [\delta h]_{nn'} = \braket{\mbf{k}, n|\delta H_{BM}|\mbf{k}, n'} 
\eea
where $\delta H_{BM}$ includes all symmetry-breaking terms, including strain and non-local tunneling, $\ket{\mbf{k},n}= c^\dag_{\mbf{k},\mbf{G}=0,n}\ket{0}, f^\dag_{\mbf{k},n} \ket{0}$ are the unperturbed basis states of the HF model. In practice, we will keep only constant terms in $\delta h^K_\eps$ and only constant and linear terms in $\delta h_\Lambda^K(\mbf{k})$, just like in the unperturbed HF model. Hence all overlaps can be computed from overlaps at $\mbf{k}=0$ (see \App{app:coeff}). To derive the form of $\delta h_\eps$, we use the fact that strain preserves $C_{2z}\mathcal{T}, P$, and transforms as a vector under $C_{3z},C_{2x}$. 
These constraints on the constant terms (see \App{app:strainHF}) yield the form 
\bea
\label{eq:strainparammain}
\delta h_\eps = \bpm
c \, \pmb{\eps} \cdot \pmb{\sigma} & c' \pmb{\eps} \cdot \pmb{\sigma}^*  & i \gamma' \eps_+ \sigma_3 \\
 &  M' \eps_+  \sigma_2 & c'' \pmb{\eps} \cdot (\sigma_0, -i \sigma_3) \\
h.c. &  & M_f \pmb{\eps} \cdot \pmb{\sigma} \\
\epm 
\eea
in terms of $\pmb{\eps} = (\eps_{xy},\eps_-)$ and the 6 strain couplings which we compute in \Tab{tab:relaxedHFparam}.

Similarly, the non-local tunneling terms obey all crystalline symmetries but commute with $P$, giving the form
\bea
\label{eq:HFrelaxperturbation}
\delta h_{\Lambda}(\mbf{k}) &= \bpm 
 \mu_1 \sigma_0 + v_2 \mbf{k} \cdot \pmb{\sigma} & v_1 \mbf{k} \cdot \pmb{\sigma}^*  &  0 \\
 & \mu_2 \sigma_0 &  v_3 \mbf{k} \cdot(\sigma_0 ,- i\sigma_3) \\
h.c. & & \mu_f \\
\epm \\
\eea
with the $P$-breaking orbital potentials $\mu_1,\mu_2$ (we always set $\mu_f = 0$) and velocities $v_1,v_2,v_3$ (see \Tab{tab:relaxedHFparam}).
We find that $v_3$ is an order of magnitude smaller and can be neglected. In fact, we will show in \Sec{sec:minimalmodels} that the velocities $v_1,v_2$ can also be neglected without losing accuracy in the flat bands. 

\begin{figure}[h]
\centering
\includegraphics[width=\columnwidth]{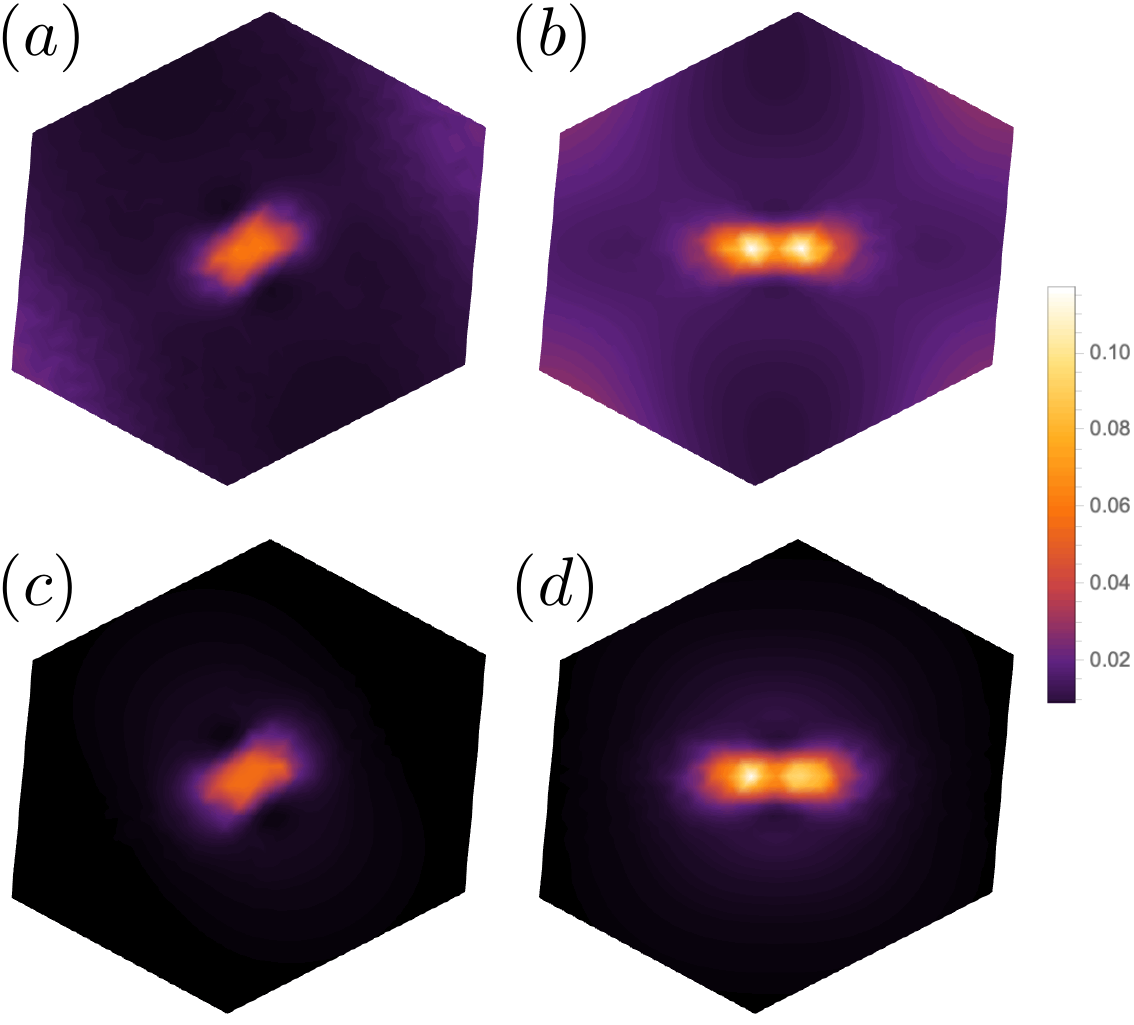}
\caption{Comparison of symmetry-breaking $\text{error}_g(\mbf{k})$. $(a)$ and $(b)$ are for the case of $C_{3z}$ in the BM and HF model, respectively. $(c)$ and $(d)$ are for the case of $C_{2x}$ in the BM and HF model, respectively at $\eps = .15\%$. 
}
\label{fig:symbreak}
\end{figure}

We compare the band structures between the BM model and generalized HF model in \Fig{BM_3_fig}, finding excellent agreement. This validates our assumption that the underlying $f$- and $c$-mode basis does not need to be altered. Rather, all symmetry breaking comes from the corrections to the Hamiltonian, and no re-Wannierization is necessary. \Fig{fig:symbreak} demonstrates that not only the spectrum, but also the symmetry-breaking patterns of the flat band eigenstates are captured by the HF model. This is quantified by the deviation of the symmetry sewing matrix from unitarity, i.e.
\bea
\label{eq:errorBg}
\text{error}_g(\mbf{k})^2 &= \sum_n(\Sigma_n[B_{g}(\mbf{k})]-1)^2
\eea
where $\Sigma_n[A]$ is the $n$th singular value of $A$ and $B_g(\mbf{k}) = U^\dag(g \mbf{k}) D[g] U(\mbf{k})$ is the sewing matrix of $g$ (here $D[g]$ is the representation of $g$ and $U(\mbf{k})$ is a column matrix of the flat band eigenvectors, the two bands closest to charge neutrality). If $g$ is a symmetry of the Hamiltonian, then $B_g(\mbf{k})$ is unitary and $\Sigma_n[B_g(\mbf{k})] = 1$. We see from \Fig{fig:symbreak} that $\text{error}(g)$ is very small away from the $\Gamma$ point. This is easily understood from the perturbation theory. The inclusion of symmetry breaking terms $\delta h$ can alter the flat bands in two ways: it can recombine the flat band eigenstates and/or mix them with dispersive band eigenstates. To begin, we consider mixing outside the flat band manifold. Recall the first order perturbation theory expression 
\bea
\delta U_0(\mbf{k}) = \sum_{n\neq0} \frac{U_n(\mbf{k}) U^\dag_n(\mbf{k})}{\Delta_n(\mbf{k})} \, \delta h(\mbf{k}) \, U_0(\mbf{k}) ,
\eea
where $U_n(\mbf{k})$ are the dispersive eigenstates with gap $\Delta_n(\mbf{k})$, and $U_0$ are the unperturbed flat bands eigenvectors. We see that away from $\Gamma$, there is a large gap $>50$meV to the closest dispersive bands compared to the $\sim 10$meV symmetry breaking terms. Thus corrections outside the flat band manifold are suppressed. Within the flat band manifold, the flat band states will be strongly mixed, but since both states are included in the sewing matrix $B_g(\mbf{k})$, only mixing outside the flat band manifold causes deviations from unitarity. Hence $\text{error}_g(\mbf{k})$ is expected to be peaked at $\Gamma$ where the gap to the dispersive bands is smallest. 

Lastly, we check numerically that including $\delta h_\Lambda$ to break particle-hole symmetry has a very weak effect on the eigenstates: the maximum deviation of the sewing matrix from unitarity is $<2.5\%$ despite visible particle-hole breaking in the spectrum \cite{2021PhRvB.103t5412S}. We will see in the following section that a full treatment of relaxation in the KV model beyond \Eq{eq:gradcoupling} results in significant violations of particle-hole symmetry in the eigenstates, which nevertheless are captured within a fully symmetry HF Hilbert space. 

\section{Fully Relaxed KV and HF Model}
\label{sec:hffullrelax}

We now consider the full effect of lattice relaxation in TBG using the Kang-Vafek model introduced in Refs. \cite{PhysRevB.107.075408,PhysRevB.107.075123}. First, it is necessary to introduce the relaxation field which arises from the classical continuum mechanics of the inter-layer interactions. 

\subsection{Atomistic Hamiltonian and Relaxation}

We index the graphene atoms in the two layers by their location $\mbf{X}_{\mbf{R},\ell, \al}$ where $\mbf{R}$ is a graphene unit cell, $\ell = \pm$ are the top and bottom layers, and $\al= $A/B are the two sublattices at positions $\pmb{\delta}_\alpha$. We decompose the positions as
\bea
\mbf{X}_{\mbf{R},\ell, \al} &= \mbf{R} + \pmb{\delta}_{\al} + \mbf{u}_{\mbf{R},\ell, \al}
\eea
where we have separated out the in-plane reference coordinates $\mbf{R}+\pmb{\delta}_\al$ from the deformation $\mbf{u}$. We follow Refs. \cite{PhysRevB.107.075408} and \cite{PhysRevB.107.075123} and consider only in-plane relaxation. 
% Note that the layer separation $\frac{1}{2}\ell d_0 \hat{z}$ can be included in $\mbf{u}$ as part of the ``relaxation" effect, where $d_0$ is the interlayer distance. We may assume $d_0$ is a constant since out-of-plane deformations only couple to the Dirac cone at second order due to the mirror symmetry of mono-layer graphene. 

In principle, the determination of the displacement field $\mbf{u}_{\mbf{R},\ell,\al}$ is a fully quantum mechanical problem solved with density functional theory (DFT) by self-consistently minimizing the electronic and lattice energies, see e.g. Refs. \cite{PhysRevB.93.235153,2023arXiv231104958J}. However, it can also be approached classically using a continuum approximation with $\mbf{r} = \mbf{R}+\pmb{\delta}_\al$:
\bea
\mbf{u}_{\ell}(\mbf{r}) \equiv \mbf{u}_{\ell,\al}(\mbf{R}+\pmb{\delta}_\al) = \mbf{u}_{\mbf{R},\ell,\al}
\eea
which assumes that $\mbf{u}(\mbf{r})$ does not rapidly vary on the atomic scale of $\pmb{\delta}_\al$. The profile of $\mbf{u}(\mbf{r})$ is variationally determined by minimizing the elasticity Hamiltonian
\bea
H[\mbf{u}_\ell(\mbf{r})] &= \sum_\ell T[\mbf{u}_\ell] + V[\mbf{u}_+-\mbf{u}_-]
\eea
where $T[\mbf{u}]$ is the kinetic energy functional depending on $\del_i u_j$, and $V[\mbf{u}_+-\mbf{u}_-]$ is the adhesion energy functional (see \App{app:relaxationcomputation}). Using the Euler-Lagrange equations for $H[\mbf{u}]$, $\mbf{u}_\ell(\mbf{r})$ can be solved numerically using an iterative procedure starting from the initial rigidly twisted configuration 
\bea
\label{eq:initialu}
\mbf{u}^0_\ell(\mbf{r}) = R(l \frac{\th}{2}) \mbf{r} - \mbf{r} 
\eea
until self-consistency is achieved. Explicit expressions can be found in \App{app:relaxationcomputation} and Ref. \cite{PhysRevB.107.075408}. The boundary conditions are periodic on the moir\'e lattice, so that $\mbf{u}_\ell(\mbf{r}+\mbf{a}_1)=\mbf{u}_\ell(\mbf{r}+\mbf{a}_2)=\mbf{u}_\ell(\mbf{r})$. Note that $\mbf{u}_\ell(\mbf{r})$ will preserve all the crystalline symmetries of the initial state, and therefore we distinguish $\mbf{u}_\ell(\mbf{r})$ from the symmetry-breaking heterostrain. 

Having determined the relaxation profile $\mbf{u}_\ell(\mbf{r})$, Ref. \cite{PhysRevB.107.075408} next obtains an effective continuum model starting from the microscopic tight-binding Hamiltonian where the hopping between two atoms only depends on their relative displacement and orientation. The microscopic Hamiltonian is
\bea
H &= \sum_{\mbf{R} \ell \al,\mbf{R}' \ell' \be} t(\mbf{X}_{\mbf{R},\ell,\al} - \mbf{X}_{\mbf{R}',\ell',\be}) c^\dag_{\mbf{R},\ell,\al} c_{\mbf{R}',\ell',\be}
\eea
and the electron operators obey discrete anti-commutation relations $\{c^\dag_{\mbf{R},\ell,\al}, c_{\mbf{R}',\ell',\be}\} = \delta_{\mbf{R}\mbf{R}'}  \delta_{\ell \ell'} \delta_{\al \be}$. The hopping function $t(\mbf{d})$ can be chosen as the Slater-Koster function for $p_z$ orbitals or fit to DFT data. By expanding $t(\mbf{X}_{\mbf{R},\ell,\al} - \mbf{X}_{\mbf{R}',\ell',\be})$ in powers of $\nabla \mbf{u}$ and $\th$, Ref. \cite{PhysRevB.107.075408} obtains a generalized continuum model taking the form
\bea
\label{eq:KVmain}
h^K_{KV}(\mbf{r}) = \bpm  v_F (- i\pmb{\nabla} - \mathcal{A}) \cdot  \pmb{\sigma}_{-\th/2}   & T(\mbf{r}) + \frac{1}{2}\{- i \pmb{\nabla}, \pmb{\Lambda}\} \\ h.c.&  v_F (- i\pmb{\nabla}+\mathcal{A}) \cdot  \pmb{\sigma}_{\th/2}  \epm  
\eea
where $\mathcal{A}(\mbf{r})$ is a relaxation gauge field computed to five shells for convergence, i.e.
\bea
\mathcal{A}_i(\mbf{r}) &= \sum_{|\mbf{G}| \leq 5 |\mbf{b}_1|} e^{i \mbf{G} \cdot \mbf{r}} \mathcal{A}_{i,\mbf{G}}, 
\eea
$T(\mbf{r}),\pmb{\Lambda}(\mbf{r})$ are inter-layer couplings computed to three shells for convergence, e.g.
\bea
T(\mbf{r}) = \sum_{i=1}^3 \Big( e^{i \mbf{q}_i \cdot \mbf{r}} T_{i}+ e^{i (-2\mbf{q}_i) \cdot \mbf{r}} T'_{i} + \sum_{j \neq i} e^{i (2\mbf{q}_i-\mbf{q}_j) \cdot \mbf{r}} T''_{i,j} \Big), 
\eea
and $\pmb{\sigma}_{\th/2} = e^{i \sigma_3 \th/2} \pmb{\sigma} e^{-i \sigma_3 \th/2}$. The gradient coupling $\pmb{\Lambda}(\mbf{r})$ in \Eq{eq:gradcoupling} is also computed to three shells. \Eq{eq:KVmain} neglects higher order gradient terms that have negligible effect. Explicit formulas for the expansion coefficients of $\mathcal{A},T,\pmb{\Lambda}$ in terms of the relaxation field $\mbf{u}(\mbf{r})$ are involved and can be found in the appendix of Ref. \cite{PhysRevB.107.075408}. We note that $v_F$ in \Eq{eq:KVmain} is computed from the full Slater-Koster parameterization to be $v_F = 5360$meV \AA, which is about $10\%$ smaller than the Dirac velocity used in the BM model (\Sec{sec:BMmainintro}). 

The band structure at $\th = 1.05^\circ$ is shown in \Fig{fig:KVmodel}, and features a larger bandwidth $\sim 15$meV, a smaller gap to the dispersive bands $\sim 5$meV, and strong particle-hole asymmetry. Despite the complexity of the Hamiltonian, we now show it is well captured by a heavy fermion representation. 

\subsection{Relaxed HF Model}

\begin{table*}[ht]
\centering
\begin{tabular}{c|cccc|cccc|cccccc}
$\theta$ \, $({}^\circ)$  & $\gamma$  & $M$ & $v$ & $v'$ & $\mu_1$ & 
$\mu_2$ & $v_1$ & $v_2$ & $c$ & $c'$ & $c''$ & $M_f$ & $\gamma'$ & $M'$ \\
\hline
\text{BM: }1.05 & -24.8 & 3.7 & -4.3 & 1.6 & 14.4 & 4.5 & 0.2 & -0.4 & -8750 & 2050 & -3362 & 4380 & -3352 & -4580 \\
 \hline 
0.95 &  -7.6 & 1.1 & -3.5 & 1. & 22.2 & 9.5 & 0.2 & -0.5 & -5160 & 480 & 4660 & -6510 & -2410 & -2560 \\
1.0 & -10.9 & -3.5 & -3.6 & 1.1 & 23.0 & 9.6 & 0.2 & -0.5 & -5750 & 550 & 4830 & -6860 & -2620 & -3020 \\
1.05 & -14.5 & -8.5 & -3.8 & 1.2 & 23.8 & 9.8 & 0.2 & -0.5 & -6290 & 630 & 4990 & -7190 & -2810 & -3430 \\
1.1 & -18.5 & -13.8 & -3.9 & 1.2 & 24.7 & 9.9 & 0.2 & -0.5 & -6760 & 710 & 5130 & -7480 & -2990 & -3790 \\
1.15 & -22.9 & -19.2 & -4.0 & 1.3 & 25.5 & 10.1 & 0.2 & -0.5 & -7180 & 790 & 5270 & -7740 & -3150 & -4110 \\
1.2 & -27.5 & -24.9 & -4.1 & 1.3 & 27.1 & 10.6 & 0.2 & -0.5 & -7540 & 880 & 5390 & -7980 & -3300 & -4400 \\
%reverse convention from notebook
 \end{tabular}
 \caption{Heavy Fermion parameters for the fully relaxed KV model. All valued reported are in meV except for the velocities $v,v',v_1,v_2$ which are in eV\,\AA. For comparison, the HF parameters in the BM model (including $\pmb{\Lambda}(\mbf{r})$) are shown in the first line. }
 \label{tab:relaxedHFparam}
\end{table*}

\begin{figure}[h]
\centering
\includegraphics[width=\columnwidth]{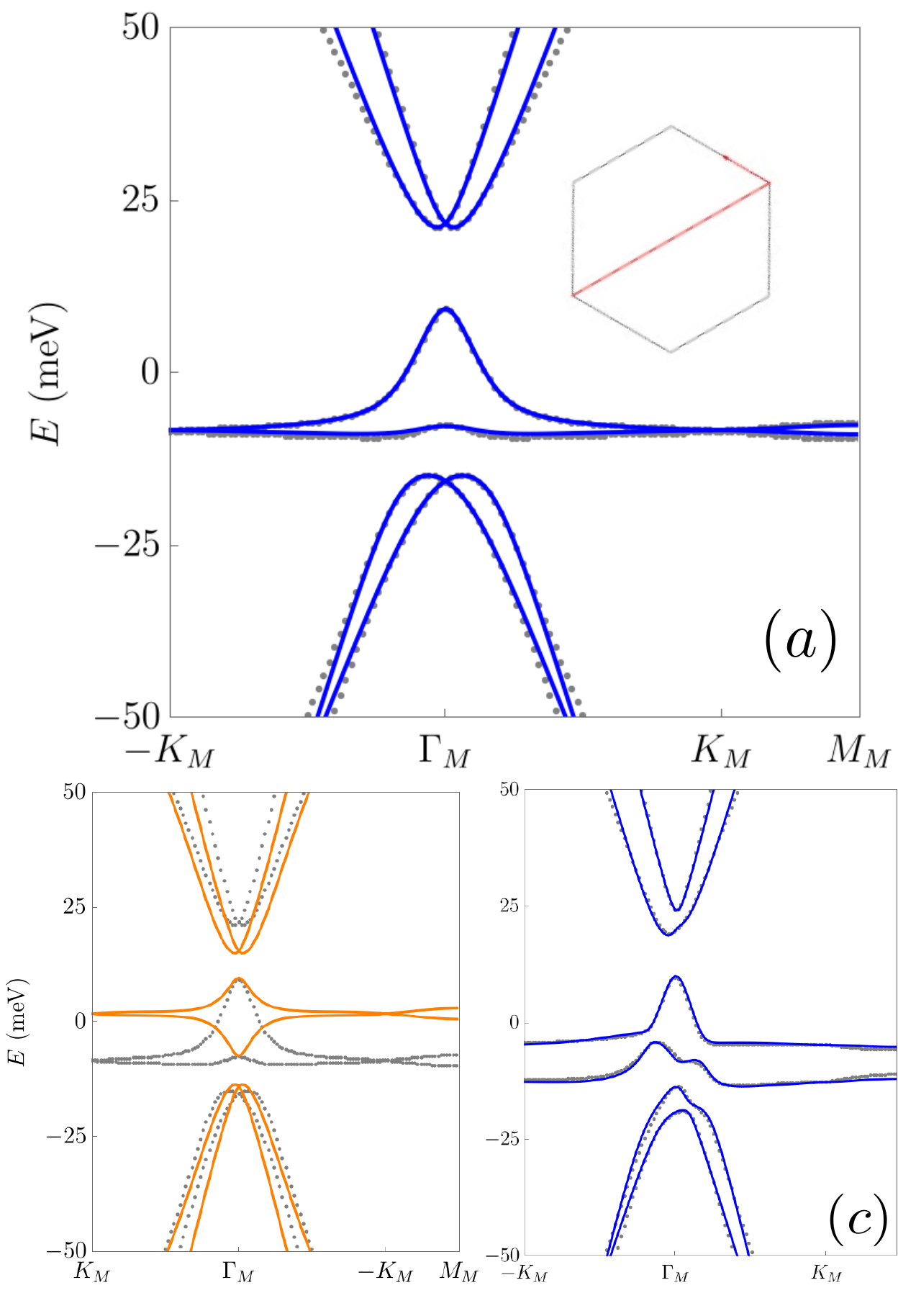}
\caption{$(a)$ Comparison of the KV model (gray dots) in \Eq{eq:KVmain} with the relaxed HF model (blue line), showing near-perfect agreement in the low-energy bands. $(b)$ The bands of the particle-hole anti-symmetrized Hamiltonian $h^-_{KV}$ are shown in orange. $(c)$ Comparison of the KV and relaxed HF model in strain, with $\eps = 0.1\%$. 
}
\label{fig:KVmodel}
\end{figure}

\begin{figure}[h]
\centering
\includegraphics[width=\columnwidth]{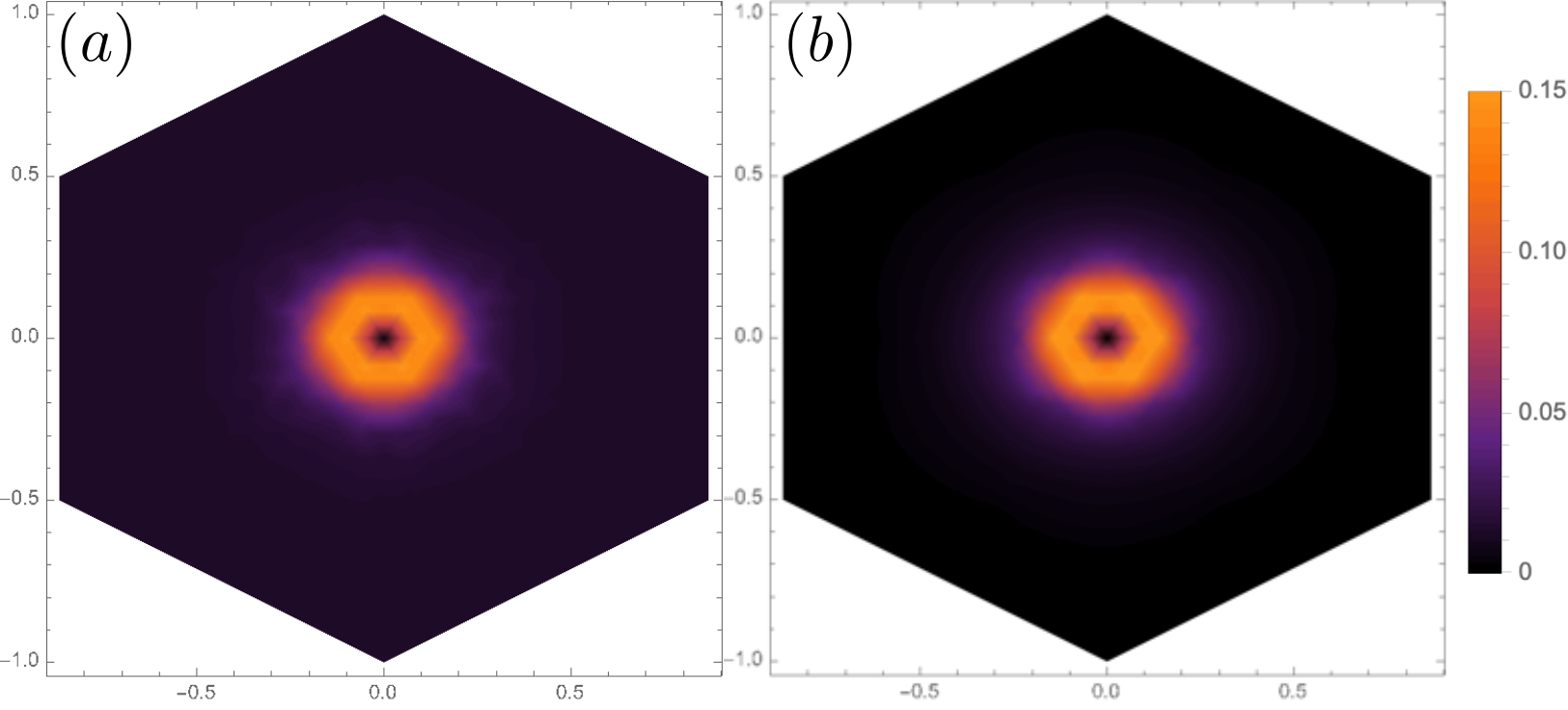}
\caption{Comparison $P$-symmetry breaking (\Eq{eq:errorBg}) in the KV model $(a)$ and the generalized HF model $(b)$.
}
\label{fig:symbreakKV}
\end{figure}

To obtain an HF model from $h^K_{KV}$, we first split up the Hamiltonian into $P$-preserving and $P$-breaking parts:
\bea
h^K_{KV} = h^+_{KV} - h^-_{KV}, \quad 2 h^\pm_{KV} = Ph^K_{KV}P \pm h^K_{KV}
\eea
where $P h^\pm_{KV} P = \pm h^\pm_{KV}$. Thus $h^-_{KV}$ preserves all of the symmetries of the original BM model (note that we have not yet included strain). Hence we can obtain the HF basis of $h^-_{KV}$ following the same exact steps as in the BM model by Wannierizing $h^-_{KV}$ (see \Fig{fig:KVmodel}b) to obtain a new set of $f$-modes, the orthogonal $c$-modes at $\Gamma$, and the new values of $\gamma, M ,v,v'$. Then $ h^+_{KV}$ is treated as a perturbation capturing the $P$-breaking due to relaxation with exactly the same form as \Eq{eq:HFrelaxperturbation}, giving new values of $\mu_1,\mu_2,v_1,v_2$. 

Finally, we consider the addition of strain to the gradient expansion formalism. In this case, \Eq{eq:initialu} is replaced by $\mbf{u}^0_\ell(\mbf{r}) = R(l \frac{\th}{2})\mbf{r} + \mathcal{E}\mbf{r} - \mbf{r}$ and the relaxation profile is solved periodically on the strained moir\'e unit cell. The resulting coefficients can be computed directly generalized the non-strained case of Kang and Vafek, and will be presented in forthcoming work \cite{KV_toappear}. Following the identical procedure as the BM model, the resulting values for $c,c',c'', M_f, \gamma', M'$ (see \Fig{fig:KVmodel}c) are computed for a range of angles and tabulated in \Tab{tab:relaxedHFparam}. 

We validate the perturbative treatment of $h^+_{KV}$ in \Fig{fig:KVmodel}a, showing nearly perfect agreement of the spectrum within $\pm 50$meV, and in \Fig{fig:symbreakKV}, showing identical patterns of $P$-breaking in the flat bands. Following the same perturbative argument as before, $\text{error}_P(\mbf{k})$ is vanishingly small away from $\Gamma$ due to the large gap between the dispersive bands and flat bands. However, unlike the breaking of the $C_3, C_{2x}$ by strain, $\text{error}_P(\mbf{k})$ is also nearly zero at $\Gamma$, but is peaked in an annulus surrounding the center of the BZ. The reason for this is a selection rule. Since $C_{3z}, C_{2x}$ are preserved, the flat band irreps $\Gamma_1,\Gamma_2$ are forbidden from mixing with the $\Gamma_3$ irreps of the nearby dispersive bands (and higher dispersive bands are over $100$meV away, proving small PH breaking in the eigenstates where $f$-modes dominate). Away from $\Gamma$, the selection rule does not hold and we thereby explain the ring-like features in \Fig{fig:symbreakKV}. 

The practical advantage of a fully symmetric Wannier basis cannot be overstated, but its true meaning is more significant. It implies that across an experimentally relevant range of strains, twist angles, and microscopic relaxation profiles,  heavy fermions provide a physical, unifying description of TBG. To substantiate this claim, we can directly compare the Wannier functions obtained in the BM model and the KV model. Since the Wannier functions in both cases obey $C_{2z}\mathcal{T},C_{3z}, C_{2x}$ and $P$ symmetry, the lowest-order Gaussian ansatz for their wavefunction $\braket{\mbf{r} \al l| f^\dag_{\mbf{R},n}|0} = w_{l \al,n}(\mbf{r})$ takes the form \cite{2022PhRvL.129d7601S}
\bea
\label{eq:Wannierapprox}
w_{l \al, 1}(\mbf{r}) &= \frac{N_\al}{\sqrt{2\pi \la_\al^2}} \lp -l \frac{x+ i y}{\la_\al}\rp^{\al-1} e^{i \frac{\pi}{4} l - |\mbf{r}|/2 \la_\al^2} \\
w_{l \al, 2}(\mbf{r}) &= \frac{N_{\bar{\al}}}{\sqrt{2\pi \la_{\bar{\al}}^2}} \lp l \frac{x - i y}{\la_{\bar{\al}}}\rp^{\bar{\al}-1} e^{-i \frac{\pi}{4} l - |\mbf{r}|/2 \la_{\bar{\al}}^2} \\
\eea
in the $K_G$ valley, where $\al = \{1,2\}, \bar{\al} = \{2,1\}$ for the A/B sublattice index. In the BM model, an overlap probability of $95\%$ (per band) is obtained with the values
\bea
(N_1,N_2) = (\sqrt{\frac{2}{3}}, - \sqrt{\frac{1}{3}}), \  (\la_1,\la_2) = (0.18, 0.23) a_M
\eea
where $a_M$ is the moir\'e lattice constant. Repeating the Wannierization on the (particle-hole symmetric part of the) KV model, we find a $95\%$ accurate fit is obtained from the similar parameters
\bea
(N_1,N_2) = (\sqrt{\frac{2}{3}}, - \sqrt{\frac{1}{3}}), \ (\la_1,\la_2) = (0.16, 0.20) a_M \ .
\eea
While the precise form of the Wannier states is dependent on the energy window taken for the dispersive bands, it is clear that the $f$-modes obtained herein reveal a hidden, fully symmetry low-energy Hilbert space in the flat bands that unites the BM and KV models. 

Finally, we will discuss the interacting parameters that define the periodic Anderson model obtained from rewriting the screened Coulomb interaction in the HF basis. Ref. \cite{2022PhRvL.129d7601S} showed in detail that the dominant interactions are an onsite Hubbard $f$-mode interaction $U_1$, a nearest-neighbor $f$-mode density interaction $U_2 \ll U_1$, the $f$-$c$ density-density interactions $W_1, W_3$ for the two $c$-mode flavors, and an $f-c$ exchange interaction $J$. We refer the reader to Ref. \cite{2022PhRvL.129d7601S} for explicit expressions, but we emphasize that the derivation of these terms depends on the symmetric $f,c$ basis obeying $C_{2z}\mathcal{T}, C_{2z},C_{2x}$ and $P$. Since we have shown in this work that the symmetric HF \emph{basis} fully describes the model with reasonable values of strain and lattice relaxation, the form of the interaction is also unchanged. In particular for the BM model, relaxation and strain do not alter the HF basis obtained in Ref. \cite{2022PhRvL.129d7601S}, and the interaction parameters are identical. In the KV model, we compute the interaction parameters from the $f$ and $c$ modes obtained with Wannier90 following the formulas in the Appendices of Ref. \cite{2022PhRvL.129d7601S} for a double-gated Coulomb interaction with screening length $\xi = 10$nm and dielectric constant $\eps = 6$. The results are shown in \Tab{tab:relaxedU}, finding quite similar values of $U_1,U_2,J$ and slightly larger values of $W_1,W_3$ between the BM model and the KV model. Note that we use the numerical Wannier functions, not the approximate forms in \Eq{eq:Wannierapprox}.

\begin{table}
\centering
\setlength{\tabcolsep}{8pt}
\begin{tabular}{c|ccccc}
$\theta$  & $U_1$ & $U_2$ & $W_1$ & $W_3$ & $J$ \\
\hline
\text{BM: }1.05 & 57.95 & 2.33 & 44.03 & 50.20 & 16.38  \\
\hline 
  0.95 & 71.3 & 1.5 & 32.3 & 46.4 & 14.4 \\
 1.0 & 72.7 & 1.7 & 36.6 & 49.4 & 15. \\
 1.05 & 72.5 & 2.1 & 41.3 & 52.7 & 15.8 \\
 1.1 & 72.1 & 2.6 & 46.2 & 56.4 & 16.8 \\
 1.15 & 71.9 & 3.1 & 51.4 & 60.4 & 17.8 \\
\end{tabular}
 \caption{Anderson Model interaction parameters for the fully relaxed KV model (see Ref. \cite{2022PhRvL.129d7601S}). All valued reported are in meV. For comparison, the HF parameters in the BM model are shown in the first line. }
 \label{tab:relaxedU}
\end{table}

\section{Analytical Expressions for the Flat Bands}
\label{app:perturbationtheory}

Having constructed the HF model and its perturbations, we are now in a position to study the flat bands and understand their behavior under strain and relaxation. We make use of a perturbation theory starting from the limit $M=0$, where the bands are perfectly flat due to an enhanced $U(4)$ symmetry \cite{2022PhRvL.129d7601S}. In this limit, $h_{HF}^K$ has the following flat band eigenvectors
\bea
\label{eq:UchiralK}
\mathcal{U}_\pm(\mbf{k}) = D_\th (0, 0, \pm v' k + \gamma, v' k \pm \gamma, -v k, \mp v k)^T 
\eea
up to a normalization, and $D_\th = \text{diag}(e^{i \th},e^{i 2\th},1,e^{i 3\th},e^{i \th},e^{i 2\th})$, $k = |\mbf{k}|$. The $SO(2)$ symmetric form of the wavefunctions arises from keeping only the $\mbf{G}=0$ shell of $c$-modes, and it is thanks to this emergent symmetry of the flat band limit of the HF model that we may find simple analytical expressions for the eigenstates. \Eq{eq:UchiralK} shows clearly how $\gamma$ is responsible for mixing the $\Gamma_1,\Gamma_2$ $c$-modes (carrying the angular momenta $0,3 = 0 \mod 3$ respectively, determined from $D_\th$) into the flat band wavefunctions near $\Gamma$, whereas at large $k$ they are dominated by the $f$-modes (with angular momenta $1,2 = \pm 1 \mod 3$) since $|v| > |v'|$. It is also straightforward to compute the (abelian) Fubini-Study metric\cite{2011EPJB...79..121R} of the two flat bands
\bea
\label{eq:FSM}
g(\mbf{k}) = \frac{1}{2}\Tr (\pmb{\nabla} \mathcal{P})^2 = \frac{4 \gamma ^2 v^2}{\left(k^2 v^2 + \gamma ^2\right)^2}
\eea
showing the characteristic $\Gamma$ point peak. In \Eq{eq:FSM}, we used $\mathcal{P} = \mathcal{U} \mathcal{U}^\dag$ and set $v' = 0$ for simplicity. It is interesting to note that $g(\mbf{k}) \to 4\pi \delta(\mbf{k})$ as $\gamma \to 0$, showing that hybridization of the $f$- and $c$-modes smooths out the intrinsic quantum geometry of the Dirac node at $\Gamma$. 

It is now straightforward to compute the corrections away from the perfectly flat band limit doing degenerate perturbation via expectation values in $\mathcal{U}_\pm(\mbf{k})$, resulting in a $2\times 2$ effective Hamiltonian for the flat bands. However, to obtain the simplest possible expressions, we will also take $v' = 0$. This approximation will still capture the key characteristics of the flat bands because $(i)$ $v' k$ vanishes near the BZ center and $(ii)$ $|v' k| < |v k|$ is subdominant for large $k$ near the BZ edge.  

We study each term in the Hamiltonian $h^K_{HF,\eps,\Lambda}(\mbf{k})$ (see \Eq{eq:Hfull}) individually. It is direct to compute
\bea
\mathcal{U}^\dag h^K_{HF} \mathcal{U} &= \frac{\gamma ^2 M}{v^2 k^2+ \gamma ^2}( \cos 3\th \sigma_3 + \sin 3\th \sigma_2)
\eea
resulting in the characteristic bump in the TBG flat bands given by $E(k) = \pm \frac{\gamma ^2 M}{v^2 k^2+ \gamma ^2}$, which is exact at $k=0$. Next we consider PH-breaking due to $\mbf{k}$-dependence in the interlayer tunneling and relaxation effects. We find
\bea
\mathcal{U}^\dag \delta h^K_\Lambda \mathcal{U} &= \frac{\gamma ^2 \mu_2}{v^2 k^2 +\gamma ^2} \sigma_0
\eea
which, including $M$ as well, yields the dispersion
\bea
E = \gamma ^2 \frac{\mu_2 \pm M}{v^2 k^2+ \gamma ^2}
\eea
so $\mu_2$ tends to flatten the bottom band and increase the dispersion in the top band since $M < 0, \mu_2 > 0$. In both the BM model with $\Lambda(\mbf{r})$ corrections and the KV model at $\th = 1.05^\circ$, $|\mu_2 + M| < 1.5$meV, which explains the very flat valence band in \Fig{BM_3_fig}(b) and \Fig{fig:KVmodel}(a) respectively. In fact, if $M + \mu_2 = 0$, a direct calculation of the determinant of the full Hamiltonian in \Eq{eq:Hfull}, $
h^K_{HF,\eps,\Lambda}(\mbf{k}) = h^K_{HF}(\mbf{k}) + \delta h^K_\Lambda(\mbf{k})$, shows that the valence band becomes perfectly flat if $M + \mu_2 = 0$. This perfect flatness relies on the truncation of the higher $\mbf{G}$ $c$-modes, but nevertheless explains strikingly $\sim 1$meV dispersion of the valence band in the fully relaxed KV model. Note that of the four parameters in $\delta h_\Lambda$, only $\mu_2$ affects the flat bands at leading order. 

Lastly, we consider the strain correction
\bea
\label{eq:strainperturb2x2}
\mathcal{U}^\dag \delta h^K_\eps \mathcal{U} &= \frac{1}{v^2k^2+\gamma^2} \Big(-2 c'' v k \gamma  (\eps_- \sin \th + \eps_{xy} \cos \th) \sigma_0  \\
&\!\!\!\!\!+ M' \gamma^2 \eps_+ (\sin 3\th \sigma_3 - \cos 3\th \sigma_2)+M_f  v^2 k^2  \pmb{\eps} \cdot \pmb{\Gamma} \Big)\\
\eea
where $\Gamma_1 = \sigma_3 \cos \th + \sigma_2 \sin \th, \Gamma_2 = -(\sigma_2 \cos \th + \sigma_3 \sin \th)$ obey $\{\Gamma_i,\Gamma_j\} = 2 \delta_{ij} \sigma_0$. A first observation is that only $c'', M',$ and $M_f$ appear at first order in perturbation theory. The most important parameter is $M_f$, the symmetry-breaking mass coupling the two $f$-modes. Away from $\Gamma$, all terms projected to the flat bands except $M_f$ go to zero and
\bea
E \to \pm M_f, \qquad (v k \gg \gamma )
\eea
showing that strain splits the flat bands around the BZ edge, where previously their degeneracy is protected by $C_3$ symmetry at the $K$ point (the BZ edge). However, the preservation of $C_{2z}\mathcal{T}$ ensures that the Dirac points cannot be gapped everywhere on the BZ within the projected model, and will be pushed away to $\mbf{k}_n$. Since $\mathcal{U}^\dag h_{HF,\Lambda,\eps}^K \mathcal{U}$ is in Dirac form, the location of the nodes can be determined from the vanishing of the $\sigma_{1,2,3}$ terms and hence is independent of $c''$. We find directly that the nodal points $\mbf{k}_n$ obey
\bea
\label{eq:diracpointsstrain}
 v^2 k^2_n &= \gamma ^2 \frac{\sqrt{M^2+M'^2\eps_+^2}}{M_f|\pmb{\eps}|},\\
 \tan \th_n &= \frac{|\pmb{\eps}|\sqrt{M^2+M'^2 \eps_+^2}+M \eps_{xy}+M' \eps_+ \eps_-}{M' \eps_+ \eps_{xy}-M \eps_-}
\eea
which break $C_{3z}$ but are mapped to each other under $\mbf{k} \to - \mbf{k}$. We see that the Dirac points  can migrate close to the $\Gamma$ point, with $k_n/|\mbf{q}_1| \sim 0.1$ for reasonable values of $\eps \sim 0.2\%$. We emphasize that the generalized HF model takes the same form for both models, and hence results of this section in \Eq{eq:diracpointsstrain} can capture the full range of flat basnds obtained from these models. 

\subsection{Minimal Models}
\label{sec:minimalmodels}

\begin{figure}[h]
\centering
\includegraphics[width=\columnwidth]{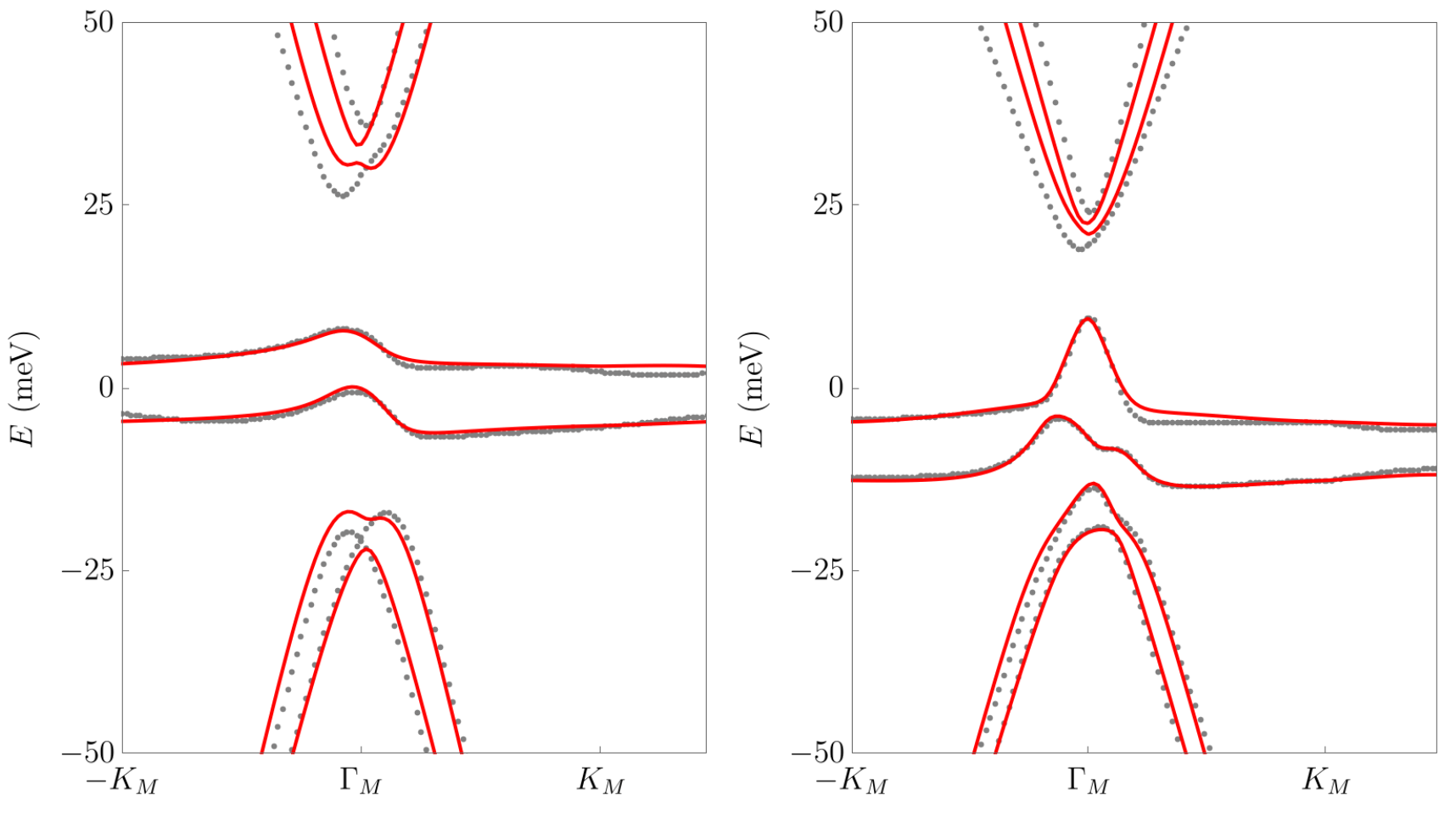}
\caption{Minimal HF model keeping  the $\mu_1,\mu_2$ PH-breaking parameters, and only the $c'',M_f$ strain parameters. With only these parameters, good quantitative agreement is still achieved in the flat bands, and the key features of the dispersive bands are retained. $(a)$ depicts the BM bands with $.15\%$ strain and PH-breaking corrections, and $(b)$ shows the KV bands with $.1\%$ strain. 
}
\label{fig:minmod}
\end{figure}

Our perturbation theory in \Sec{app:perturbationtheory} has shown that the flat bands are determined by only a few HF parameters. Since the flat bands dominate the low-energy Hilbert space (and comprise it entirely in projected calculations), we will write down a minimal HF model with the fewest possible parameters required to capture the flat bands. 

Of the PH-breaking parameters $\mu_1, \mu_2, v_1,v_2$, only $\mu_2$ is required to describe the flat band dispersion in perturbation theory (\Sec{app:perturbationtheory}), while $\mu_1$ sets the gap to the conduction bands. Hence, as a minimal model of relaxation, we keep only $\mu_1,\mu_2$. Of the strain parameters $c,c',c'',\gamma',M',M_f$, only $c'',M'$ and $M_f$ appear in perturbation theory via \Eq{eq:strainperturb2x2}. Furthermore, since $\eps_+$ is smaller than the $\eps_-$ due to the nonzero Poisson ratio of graphene $\nu =0.16$, we can even neglect $M'$ (see \Eq{eq:strainparammain}). Thus, we keep only $M_f$ and $c''$ in the minimal model. 

The results are shown in \Fig{fig:minmod} for both the extended BM model and the KV model, which demonstrates the excellent agreement retained by the minimal model within the flat bands. The features of the dispersive bands are also qualitatively accurate, and energy differences between the minimal HF model and full continuum model are confined to less than $3$meV at low energies. This agreement is rather good given the enormous reduction in parameters, and showcases the ability of heavy fermions to efficiently and accurately model the realistic electronic structure of TBG. 

\section{Conclusion}
\label{sec:conclusion}

This work has constructed a single-particle heavy fermion model for TBG in the presence of realistic strain and lattice relaxation in two models: the perturbed BM model with particle-hole breaking and minimally-coupled heterostrain, and the KV model incorporating fully microscopic relaxation. Despite the significant differences between these Hamiltonians, they are described by the same (generalized) heavy fermion model, albeit with different parameterizations. This model accurately reproduces the lowest 6 bands within a $\pm 50$meV window, as well as reproducing the patterns of symmetry breaking in the eigenstates. All parameters of the model are computed directly from the $f$-mode wavefunctions obtained with Wannier90 and the continuum Hamiltonians --- no fitting or optimization is performed. Although we have obtained the heavy fermion parameters numerically, good analytical estimates are possible \cite{2023LTP....49..640C} and left for other work. 

We made use of the simple form of the HF model to analytically obtain flat band wavefunctions and their dispersions in the presence of relaxation and strain. We found that the key features of the active bands are determined by a small number of parameters, the three symmetry-breaking terms $\mu_2, M',$ and $M_f$ in addition to the  $v,\gamma,M$ terms of the original HF model, which could be useful for the study of reduced models. 

Ultimately, the purpose of the HF model is not merely to simplify the single-particle physics, but to approach the strongly interacting problem. It is worth repeating that a minimal tight-binding model within a reduced Hilbert space --- the typical tool in this pursuit --- is unavailable due to the stable topological obstruction protected by the $P C_{2z} \mathcal{T}$ symmetry emergent in the BM model. Even $P$-breaking perturbations do not provide an easy escape since these perturbations reduce the gap to the dispersive bands, making a projection onto the flat bands unreliable due to the large Hubbard $U_1$. 

The HF model was originally proposed to circumvent these difficulties \cite{2022PhRvL.129d7601S}. It provides a faithful model whose simple form is amenable to analytical calculation, and whose locality and low dimensionality (after truncating the $c$-modes) make it suitable for intensive numerics. The inclusion of realistic strain and relaxation poses no difficulty for the formalism, underscoring the advantage of heavy fermions. In Ref. \cite{2025arXiv250208700H}, we leverage this result to develop an analytical understanding of the IKS as a heavy fermion state. 

\section{Acknowledgments}
The authors are grateful for important conversations and collaborations with Yves Kwan, Antoine Georges, Xi Dai, Andy Millis, Francisco Guinea, Roser Valent\'i, Giorgio Sangiovanni, Tim Wehling, Qimiao Si, Jian Kang, Dmitri K. Efetov, Keshav Singh, Liam Lau, Daniel Kaplan, Ryan Lee, Gautam Rai, Lorenzo Crippa, and Michael Scheer. B. A. B. was supported by Office of Basic Energy Sciences, Material Sciences and Engineering Division, U.S. Department of Energy (DOE) under Contracts No. DE-SC0016239 and the Simons Collaboration on New Frontiers in Superconductivity (SFI-MPS-NFS-00006741-01). J. H.-A. is supported by a Hertz Fellowship, with additional support from DOE Grant No. DE-SC0016239. J. Y. is supported by the Gordon and Betty Moore Foundation through Grant No. GBMF8685 towards the Princeton theory program and through the Gordon and Betty Moore Foundation’s EPiQS Initiative (Grant No. GBMF11070). D.C. acknowledges the hospitality of the Donostia International Physics Center, at which this work was carried out, and acknowledges support by the Simons Investigator Grant No. 404513, the Gordon and Betty Moore Foundation through Grant No. GBMF8685 towards the Princeton theory program, the Gordon and Betty Moore Foundation’s EPiQS Initiative (Grant No. GBMF11070), Office of Naval Research (ONR Grant No. N00014-20-1-2303), Global Collaborative Network Grant at Princeton University, BSF Israel US foundation No. 2018226. H.H. was supported by the European Research Council (ERC) under the European Union’s Horizon 2020 research and innovation program (Grant Agreement No. 101020833).

\bibliography{refs}
\bibliographystyle{apsrev4-1} 

\appendix
\onecolumngrid

\section{Single-Particle Hamiltonians}

In this Appendix, we briefly review our conventions for the original Bistrizter-MacDonald (BM) model and its symmetries (\App{app:OGBM}) before introducing strain  and particle-hole breaking terms (\App{eq:strainBM}) into the Hamiltonian.

\subsection{Original BM Model}
\label{app:OGBM}

The Hamiltonian of twisted bilayer graphene is built from the low-energy Dirac cones of the top $(l= +)$ and bottom $(l=-)$ layers of graphene coupled by an interlayer tunneling term. Without twisting, each layer hosts two Dirac cones at $\mbf{K}_G = \frac{4\pi}{3a}(1,0)$ and $\mbf{K}' = -\mbf{K}_G$, corresponding to the graphene lattice vectors $\mbf{a}_{G,1} = a(1,0), \mbf{a}_{G,2} = C_{3z} \mbf{a}_1$. Here $a =0.246$nm is the graphene lattice constant and $C_{3z} = R(2\pi/3)$ with $R(\th) = e^{-i \th \sigma_2}$ the rotation matrix on 2D vectors. In our convention, the $l = \pm$ layers of graphene are rotated by $\pm \th/2$ where $\th$ is the relative twist angle near $1^\circ$. For small angles, the $\mbf{K}_G$ and $\mbf{K}'_G$ valleys are decoupled. We will consider the $K$ valley and obtain the $K'$ valley by time-reversal. Rotating the layers displaces the $K$-valley Dirac cones by
\bea
\label{eq:BMq}
\mbf{q}_1 &= R(-\th/2) \mbf{K}_G - R(\th/2) \mbf{K}_G = -k_\th (0,1), \quad \mbf{q}_{j+1} = C_{3z} \mbf{q}_j, \quad k_\th = \frac{4\pi}{3a}  \times 2 \sin \frac{\th}{2} \ . \\
\eea
The moir\'e reciprocal lattice is generated by $\mbf{b}_i  = \mbf{q}_3 - \mbf{q}$, and leads to a momentum space lattice defined by $\mbf{Q} \in \mathds{Z} \mbf{b}_1 + \mathds{Z} \mbf{b}_2 \pm \mbf{q}_1$ with the $\pm$ corresponding to the Dirac cone in layer $l = \pm$. We use a convention for the untwisted graphene Dirac cone where $h_G(\mbf{K}_G+\mbf{k}) =  v_F \mbf{k} \cdot \pmb{\sigma}+\dots$, and $\pmb{\sigma} = (\sigma_1, \sigma_2)$ is the vector of Pauli matrices acting on the graphene sublattice index $\al = A,B$. The single-particle Hamiltonian can then be written in terms of its matrix elements on the tensor product of graphene sublattice $\al$ and momentum space $\mbf{Q}$ lattice as (see Ref. \cite{2018arXiv180710676S}, which we follow closely)
\bea
\label{eq:hBM}
\null [h^K_{BM}]_{\mbf{Q} \al,\mbf{Q}' \al'}(\mbf{k}) =  v_F  \delta_{\mbf{Q},\mbf{Q}'} (\mbf{k} - \mbf{Q}) \cdot \pmb{\sigma}_{\al \al'} + \lp \sum_{i =1}^3  \delta_{\mbf{Q},\mbf{Q}'+\mbf{q}_j}  [T_j]_{\al \al'}+ h.c. \rp, \quad   v_F k_\th = 10.12 \text{eV} \times 2 \sin \frac{\th}{2} 
\eea
where the term in parentheses is the Bistritzer-MacDonald inter-layer tunneling
\bea
T_{j+1} = w_0 \sigma_0 + w_1(\sigma_1 \cos \frac{2\pi}{3}j + \sigma_2 \sin \frac{2\pi}{3}j), \qquad w_1 = 110 \text{meV}, \ w_0 =0.8 w_1 \ .
\eea
The momentum $\mbf{k} = k_1 \mbf{b}_1+ k_2 \mbf{b}_2$ is defined in the moir\'e Brillouin zone spanned by $k_i \in [0,1)$. For numerics, we take a large circular cutoff on the $\mbf{Q}$ lattice and define the $\Gamma$ point to be $\mbf{k} = (0,0)$. The Hamiltonian in the $K'$ valley is given by
\bea
h^{K'}_{\mbf{Q} \al,\mbf{Q}' \al'}(\mbf{k}) &= h^K_{-\mbf{Q} \al,-\mbf{Q}' \al'}(-\mbf{k})^* 
\eea
using spin-less time-reversal symmetry. The full $SU(2)$ spin symmetry is obtained by taking two copies of this model for spin $\uparrow$ and spin $\d$. 

The symmetries of the BM model play an important role. The intra-valley crystallographic symmetries of the model obey
\bea
D[g] h(\mbf{k}) D^{-1}[g] &= h(g \mbf{k})
\eea
forming the magnetic space group $G = p 6'2'2$ are generated by $C_{3z}, C_{2z} \mathcal{T}, C_{2x}$ with the representations ($K$ denotes complex conjugation)
\bea
\label{eq:BMsym}
D[C_{3z}]_{\mbf{Q} \al,\mbf{Q}' \al'} &= \delta_{\mbf{Q},C_{3z} \mbf{Q}'} [e^{i \frac{2\pi}{3} \sigma_3}]_{\al \al'} \\
D[C_{2z}\mathcal{T}]_{\mbf{Q} \al,\mbf{Q}' \al'} &= \delta_{\mbf{Q},\mbf{Q}'} [\sigma_1]_{\al \al'} K \\
D[C_{2x}]_{\mbf{Q} \al,\mbf{Q}' \al'} &= \delta_{\mbf{Q},C_{2x}\mbf{Q}'} [\sigma_1]_{\al \al'}  \ .
\eea
The model in \Eq{eq:hBM} neglects the $O(\th)$ opposite rotations of $\mbf{k}$ within each layer and as such enjoys a \emph{unitary} anti-commuting operator $P$ obeying
\bea
\label{eq:PHBMsym}
D[P] h(\mbf{k}) D^\dag[P] &= -h(-\mbf{k}), \qquad D[P] = \zeta_\mbf{Q} \delta_{\mbf{Q},-\mbf{Q}'} \delta_{\al \al'}, \qquad \zeta_{m\mbf{b}_1+n \mbf{b}_2 \pm \mbf{q}_1} = \pm 1 \ .
\eea
We refer to $P$ as a particle-hole symmetry because it relates states of energies $\pm E$ and $\pm \mbf{k}$. 

\subsection{Graphene with Strain}
\label{app:strain}

In order to be self-contained, we derive the strain pseudo gauge field arising from deforming monolayer graphene \cite{2022PhRvB.105x5408N}. We choose the graphene lattice vectors to be
\bea
\mbf{a}_{G,1} = a(1,0), \quad \mbf{a}_{G,2} = C_3 \mbf{a}_{G,1}
\eea
and the orbital sites to be $\mbf{r}_{1} = (\mbf{a}_{G,1} - \mbf{a}_{G,2})/3, \mbf{r}_2 = C_6  \mbf{r}_{1}$. The $K_G$-point is located at $\mbf{K} = \frac{4\pi}{3 a}(1,0)$. The nearest neighbor tight-binding model without strain is
\bea
h(\mbf{k}) &= \bpm 0 & t^*(\mbf{k}) \\ t(\mbf{k}) & 0 \epm, \qquad  t(\mbf{k}) &= \sum_{j=0}^2 t[\pmb{\delta}_j] e^{- i \mbf{k} \cdot \pmb{\delta}_j}, \qquad \pmb{\delta}_j = C_3^j(\mbf{r}_2 - \mbf{r}_1)
\eea
where $\pmb{\delta}_j$ are the nearest-neighbor distances and $t[\mbf{d}]$ is the hopping strength between two carbon $p_z$ orbitals a distance $\mbf{d}$ apart. To derive the strain pseudo-gauge field used to couple strain to the BM model (note that full strain coupling in the KV model is beyond this approximation), we assume that the hopping function obeys $t[\mbf{d}] = t[|\mbf{d}|]$ and set $t[\pmb{\delta}_j] = t[|\pmb{\delta}_j|] = t$. Around the $K_G$ point, we find
\bea
h(\mbf{K} + \mbf{k}) &= v_F \, \mbf{k} \cdot \pmb{\sigma} + \dots, \qquad v_F = -\frac{\sqrt{3}}{2} t a
\eea
and we note that $v_F > 0$ since $t<0$. 

Now we introduce uniform strain by taking
\bea
\label{eq:hstrainedfull}
h(\mbf{k}) \to \tilde{h}(\mbf{k}) &= \bpm 0 & \tilde{t}^*(\mbf{k}) \\ \tilde{t}(\mbf{k}) & 0 \epm, \qquad  \tilde{t}(\mbf{k}) &= \sum_{j=0}^2 t[\tilde{\pmb{\delta}}_j] e^{- i \mbf{k} \cdot \tilde{\pmb{\delta}}_j}, \qquad \tilde{\pmb{\delta}}_j = (1+\mathcal{E}) C_3^j(\mbf{r}_2 - \mbf{r}_1) = \pmb{\delta}_j + \mathcal{E} \pmb{\delta}_j 
\eea
since strain transforms any orbital location $\mbf{r} \to (1+\mathcal{E})\mbf{r}$ at leading order. Taylor expanding the hopping to leading order, we find 
\bea
t[\tilde{\pmb{\delta}}_j] = t + \pmb{\nabla} t \cdot \mathcal{E} \pmb{\delta}_j = t (1 - \be \, \hat{\pmb{\delta}}_j \cdot \mathcal{E} \hat{\pmb{\delta}}_j ), \qquad \be = -\left. \frac{\del \log t(r)}{\del \log r} \right|_{r = |\pmb{\delta}_j|}, \qquad \hat{\pmb{\delta}_j} = \frac{\pmb{\delta}_j}{|\pmb{\delta}_j|} 
\eea
where $\be > 0$ functions as the strain coupling. Next we note that under $\mbf{a}_i \to (1 + \mathcal{E})\mbf{a}_i$, we have $\mbf{b}_i \to (1- \mathcal{E}^T) \mbf{b}_i$ to satisfy $\mbf{a}_i \cdot \mbf{b}_j = \delta_{ij}$, and we define the strained $\mbf{K}$ point as $\tilde{\mbf{K}} = (1 -\mathcal{E}^T)\mbf{K}$. The above formula now gives
\bea
\tilde{h}(\tilde{\mbf{K}}+\mbf{k}) = v_F (\mbf{k} - \mbf{A}[\mathcal{E}]) \cdot \pmb{\sigma} + \dots, \qquad \mbf{A}[\mathcal{E}] = \frac{\sqrt{3} \be}{a} (\frac{\mathcal{E}_{11}-\mathcal{E}_{22}}{2},-\frac{\mathcal{E}_{12}+\mathcal{E}_{21}}{2})
\eea
using perturbation theory on \Eq{eq:hstrainedfull} to first order in $\mbf{k}$ and $\mathcal{E}$.

\subsection{BM Model with Heterostrain}
\label{eq:strainBM}

In this section, we consider the addition of strain, which reduces the symmetry group, increases the bandwidth, and decreases the gap to the dispersive bands. 

We follow closely the derivation in Ref. \cite{2022PhRvB.105x5408N} for the introduction of strain into the BM model. There are two main effects. First, strain deforms the moir\'e $\mbf{q}$ vectors, breaking $C_{3z}$ symmetry and altering the $\mbf{Q}$ momenta that enter in the kinetic term. Second, strain couples directly to the Dirac cones in the form of a pseudo-vector potential $\mbf{k} \to \mbf{k} - \mbf{A}$ as in monolayer graphene. The introduction of strain in the KV model is more involved, since the relaxation profile of the lattice is also perturbed \cite{KV_toappear}. 

We start by implementing strain at the graphene scale. A uniform strain transforms of the lattice sites $\mbf{r} \to (1+\mathcal{E})\mbf{r}$ where $\mathcal{E}$ is a general $2\times 2$ matrix that we parameterize by in layer $l$ by
\bea
\mathcal{E}_l = \bpm \eps^l_{11} & \eps^l_{12} \\ \eps^l_{21} & \eps^l_{22} \epm \equiv \bpm \eps^l_{+}+\eps^l_{-} & \eps^l_{xy} + \Omega^l \\ \eps^l_{xy} - \Omega^l & \eps^l_{+}-\eps^l_{-} \epm
\eea
whose meanings are: $\eps_{\pm}$ is the isotropic/anisotropic strain, $\eps_{xy}$ is the shear strain, and $\Omega$ is the rigid rotation which defines the twist angle. To leading order, meaning that we neglect $\mathcal{E} R(\th/2) $ since it is second order, the strained lattice vectors in the top and bottom layers are
\bea
\label{eq:strainedaG}
\mbf{\tilde{a}}^l_{G,i} &= \big( R(l \th/2) + \mathcal{E}_l \big) \mbf{a}_{G,i} + O(\eps \th)\ .
\eea
In the BM model, the moir\'e BZ is determined by the \emph{differences} of the momentum vectors in the two layers $l = \pm$. Hence we define
\bea
\mathcal{E} = \mathcal{E}_{-} - \mathcal{E}_{+}, \qquad \eps_{\pm} = \eps^{-}_{\pm} - \eps^{+}_{\pm}, \qquad \eps_{xy} = \eps^{-}_{xy} - \eps^{+}_{xy}, \qquad \Omega = \Omega^{-} - \Omega^{+} \ .
\eea
The typical values of $\eps_{\pm}, \eps_{xy}$ are $.001 -0.004$ and are less than $1\%$. To illustrate the effects of strain, we choose to apply a compressive uniaxial strain along the $\hat{x}$ axis with strength parameterized by $\eps^l$. Graphene is a conventional elastic material where compression along one direction leads to elongation along the other determined by Poisson's ratio $\nu_G =0.16$. Thus the strain tensor in layer $l$ is
\bea
\label{eq:strainsym}
\mathcal{E}^l = \bpm - \eps^l & 0 \\ 0 & \nu_G \eps^l \epm, \qquad \eps^l_+ = \frac{(\nu_G-1) \eps^l}{2}, \quad \eps^l_- = -\frac{(\nu_G+1) \eps^l}{2}, \quad  \eps^l_{xy} = 0  \ .
\eea
Rotating the strain axis by an angle $\varphi$ is accomplished by taking $\mathcal{E}^l \to R^{-1}(\varphi) \mathcal{E}^l  R(\varphi)$, from which we find that  $(\eps_{-}, \eps_{xy})^T \to R(2\varphi) (\eps_{-}, \eps_{xy})^T$ transforms as a vector. For simplicity, we take $\varphi = 0$ unless otherwise stated. Secondly, we also only consider heterostrains (opposite strain in each layer) such that $\mathcal{E}^{-} = - \mathcal{E}^{+} \equiv \mathcal{E}/2$. Homostrain (same strain in each layer) can be neglected to leading order as we will show momentarily. 

We now write down the BM model in the presence of strain. To derive the deformed momentum space latitce, we first note that the graphene reciprocal lattice vectors $\mbf{b}_{G,i}$ obey $\mbf{b}_{G,i} \cdot \mbf{a}_{G,j} = 2\pi \delta_{ij}$. Adding strain, \Eq{eq:strainedaG} shows that
\bea
\mbf{\tilde{b}}_{G,i}^l &= \big( (R(l \th/2) + \mathcal{E}_l)^T \big)^{-1} \mbf{b}_{G,i} = \big( R(l \th/2) - \mathcal{E}^T_l \big) \mbf{b}_{G,i} + O(\eps \th)
\eea
since $R(\th)^T = R(\th)^{-1}$. The $K$ point in the the corresponding strained and rotated BZ is 
\bea
\mbf{\tilde{K}}^l & =  \big( R(l \th/2) - \mathcal{E}_l^{T} \big) \mbf{K}_G, \quad \mbf{K}_G = \frac{4\pi}{3a}(1,0), \quad a =0.246 \text{nm}
\eea
and thus the moir\'e $\mbf{q}_j$ vectors become (see \Eq{eq:BMq})
\bea
\mbf{\tilde{q}}_j &=  \big( R(-\th/2) - R(\th/2) \big) \mbf{K}_j - \mathcal{E}^T \mbf{K}_j, \qquad \mbf{K}_j = C^{j-1}_3 \mbf{K}_G \\
\eea
where we separated out the first (unstrained) term from the strain correction. The moir\'e reciprocal lattice is finally defined by
\bea
\label{eq:btildestrain}
\mbf{\tilde{b}}_i &= \mbf{\tilde{q}}_3 - \mbf{\tilde{q}}_i = \mbf{b}_i  - \mathcal{E}^T (\mbf{K}_3 - \mbf{K}_i) \\
\eea
Since the reciprocal lattice vectors are deformed from the unstrained case, the deformed $\tilde{\mbf{Q}}$ momentum space lattice alters the kinetic term. This is one mechanism through which strain changes the spectrum. However, strain also couples directly to the Hamiltonian in the form of a strain pseudo-vector potential which shifts the Dirac points oppositely in opposite layers. Using \App{app:strain} gives
\bea
\label{eq:strainA}
h_{K,l}(\mbf{k}) &=  v_F  (R(l \th/2) + \mathcal{E}^l)^{-1} (\mbf{k} - \mbf{A}^l ) \cdot \pmb{\sigma}, \qquad \mbf{A}^l  = \frac{\be}{a/\sqrt{3}}  (\eps_-^l , -\eps_{xy}^l) \\
&\approx  v_F (\mbf{k} -  \mbf{A}^l) \cdot \pmb{\sigma}
\eea
where $\be \approx 3.14$ measures the change of the hopping $t_0$ as the lattice vector is varied (due to strain)\cite{2022PhRvB.105x5408N}. We dropped the $O(\th), O(\eps)$ terms in the second line of \Eq{eq:strainA} since they lead to very small corrections subdominant to the strain vector potential. In this limit, homostrain $\mbf{A}^l = \mbf{A}$ is a flat gauge transformation and simply shifts the origin of $\mbf{k}$ identically in both layers. For this reason, we can focus on heterostrain. To summarize, the BM model in the presence of uniaxial heterostrain is
\bea
\label{eq:hBMstrain}
\tilde{h}^K_{\mbf{\tilde{Q}} \al,\mbf{\tilde{Q}}' \al'}(\mbf{k}) = \delta_{\mbf{\tilde{Q}},\mbf{\tilde{Q}}'} (v_F  (\mbf{k}- \mbf{\tilde{Q}}) + \frac{\be}{a/\sqrt{3}} \zeta_\mbf{\tilde{Q}} (\eps_-/2,- \eps_{xy}/2) ) \cdot \pmb{\sigma}_{\al \al'} + \lp \sum_{i =1}^3  \delta_{\mbf{\tilde{Q}},\mbf{\tilde{Q}}'+\mbf{\tilde{q}}_j}  [T_j]_{\al \al'}+ h.c. \rp \ .
\eea
Note that only the kinetic Dirac-like term depends explicitly on the strain through the appearence of $\mbf{k}- \mbf{\tilde{Q}}$, whereas the potential term is still a nearest neighbor hopping on the momentum space lattice. Note that $\mbf{k}$ ranges over the strained BZ (see \Eq{eq:btildestrain}), defined by
\bea
\{k_1 \mbf{\tilde{b}}_1 + k_2 \mbf{\tilde{b}}_2| k_i \in (-\pi,\pi) \} = \{k_1 \mbf{b}_1 + k_2 \mbf{b}_2 - k_1 \mathcal{E}^T(\mbf{K}_3 - \mbf{K}_1)- k_2 \mathcal{E}^T(\mbf{K}_3 - \mbf{K}_2) | k_i \in (-\pi,\pi) \}
\eea
which defines a one-to-one map between the unstrained (moir\'e) BZ and the strained (moir\'e) BZ
\bea
\mbf{k} \to \mbf{k} - \sum_i (\mbf{k} \cdot \mbf{a}_i ) \mathcal{E}^T(\mbf{K}_3 - \mbf{K}_i), \qquad (\mbf{a}_i \cdot \mbf{b}_j = \delta_{ij})  \ .
\eea
The Main Text shows example band structures in the $K$ valley which demonstrate the strong effect of strain on the flat bands. Note that strain preserves time-reversal, and the psuedo-vector potential is opposite in opposite valleys. 

We now discuss the symmetries of $\tilde{h}(\mbf{k})$ (\Eq{eq:hBMstrain}) in the presence of strain. A general uniaxial strain deforms the lattice and breaks $C_3$ and $C_{2x}$ but importantly preserves both $C_{2z} \mathcal{T}$ and $P$, which follows respectively from $\sigma_1 \mbf{A} \cdot \pmb{\sigma} \sigma_1 = \mbf{A} \cdot \pmb{\sigma}^*$ (see \Eq{eq:BMsym}) and $\mbf{A}^{l} = - \mbf{A}^{-l}$ (see \Eq{eq:PHBMsym}). The fact that $P$ is still an anti-commuting operator depends on choosing a pure heterostrain as well as dropping the $O(\th), O (\eps)$ terms that rotate the the Dirac cones differently in the top and bottom layers. Both are excellent approximations, showing that $P$ is a good symmetry of the low energy physics. 

What is the consequence of these symmetries? $C_{2z} \mathcal{T}$ is sufficient to protect fragile topology through the Euler number, and $PC_{2z}\mathcal{T}$ protects a stable topological anomaly that obstructs a lattice model at all energies. Together, these symmetries obstruct any local basis of the flat bands in a finite lattice model. The heavy fermion model, to be introduced in the following section, evades this obstruction using the continuum conduction electrons.

\section{Heavy Fermion Hamiltonian and Symmetries}
\label{app:HFreview}

 We briefly review the Heavy Fermion mapping of TBG from \cite{2022PhRvL.129d7601S}. The two nearly flat bands of TBG are anomalous in the presence of $C_{2z}\mathcal{T}$ and $P$, meaning that no lattice model can reproduce their wavefunctions. In fact, this anomaly cannot be lifted by including any particle-hole symmetric pair of bands in the lattice model. However, a fully symmetric low-energy model can be obtained in a different basis: the heavy fermion basis. It consists of two flavors of Dirac electrons (one carrying the 2D $\Gamma_3$ irrep, one carrying the 2D $\Gamma_1 \oplus \Gamma_2$ rep) and a pair of localized Wannier states forming $p_x$-$p_y$ orbitals at the moir\'e unit cell center (carrying the $E_{1a}$ irrep). 
 
The BM model can be projected into this basis and the following the single-particle Hamiltonian in the $K$ valley is obtained (the basis is ordered $\Gamma_3, \Gamma_1 + \Gamma_2, E_{1a}$)
\bea
h_K(\mbf{k}) &= \bpm 
 & v(k_x \sigma_0 + i k_y \sigma_3) &  \gamma \sigma_0 + v'(k_x \sigma_1 + k_y \sigma_2 )\\
v(k_x \sigma_0 - i k_y \sigma_3) & M \sigma_1 & \\
 \gamma \sigma_0 + v'(k_x \sigma_1 + k_y \sigma_2 ) & & 0 \\
\epm
\eea
neglecting higher shells of the Dirac electrons for simplicity. The parameters of the model are given in \Tab{tab:relaxedHFparam}. The intra-valley symmetries of this Hamiltonian are
\bea
D[C_{3z}] &= \bpm e^{i \frac{2\pi}{3} \sigma_3} & & \\
& \sigma_0 & \\ & & e^{i \frac{2\pi}{3} \sigma_3} \epm, \qquad D[C_{2z}\mathcal{T}] = \bpm \sigma_1 & & \\
& \sigma_1 & \\ & & \sigma_1 \epm K \\
D[C_{2x}] &= \bpm \sigma_1 & & \\
& \sigma_1 & \\ & &  \sigma_1\epm, \qquad D[P] = -i \bpm \sigma_3 & & \\
& \sigma_3 & \\ & & -\sigma_3 \epm \\
\eea
and the Hamiltonian in the $K'$ valley can again be obtained by time-reversal, and spin $SU(2)$ symmetry is implemented by taking two copies of the $K$ and $K'$ Hamiltonians, leading to a four-fold degeneracy overall. This degeneracy is due to the $U(1)$ charge, $U(1)$ valley, and $SU(2)$ spin symmetries which yield a $U(2) \times U(2)$ symmetry group, the factors corresponding to the two valleys. For clarity, we write the full model
\bea
\label{eq:hHF}
h_{HF}(\mbf{k}) &=  \bpm 
 & v(k_x \tau_3 \sigma_0 + i k_y \tau_0 \sigma_3) &  \gamma \tau_0 \sigma_0 + v'(k_x \tau_3 \sigma_1 + k_y \tau_0 \sigma_2 )\\
v(k_x \tau_3 \sigma_0 - i k_y \tau_0 \sigma_3) & M \tau_0 \sigma_1 & \\
 \gamma \tau_0 \sigma_0 + v'(k_x \tau_3 \sigma_1 + k_y \tau_0 \sigma_2 ) & & 0 \\
\epm \otimes s_0
\eea
where $\tau_i$ and $s_i$ are sets of Pauli matrices representing valley and spin respectively. The inter-valley crystallographic symmetries are spin-less time-reversal and two-fold rotation
\bea
D[\mathcal{T}] &= \tau_1 K, \quad D[C_{2z}] = \tau_1 \sigma_1
\eea
which both take $\mbf{k} \to - \mbf{k}$. The global symmetries are generated by the Hermitian commuting charge $\tau_3$ (valley) and $s_i$ (spin), forming $U(2)\times U(2)$. We will discuss the enhanced symmetries of the flat bands in coming work \cite{comingwork}.

\subsection{Relaxation and Strain}
\label{app:strainHF}
\label{app:coeff}

We now discuss the topological heavy fermion model characterizing the strained case. Note that the small strains we consider can be up to 20 meV, which has a large effect on the flat bands but still is small compared to the overall energy window within the HF Hilbert space.

We first need to parameterize all the mass terms that are now allowed in the model. Pure heterostrain preserves $C_{2z} \mathcal{T}$ and $P$ per the discussion after \Eq{eq:strainA}. Recall that the representations of these symmetries are
\bea
D[C_{2z}\mathcal{T}] = \bpm \sigma_1 & & \\
& \sigma_1 & \\ & & \sigma_1 \epm K, \qquad D[P] = -i \bpm \sigma_3 & & \\
& \sigma_3 & \\ & & -\sigma_3 \epm \ . \\
\eea
We enumerate all the new mass terms allowed by these symmetries alone (e.g. $M$ and $\gamma$ are allowed but already appear in the model). Only $\sigma_1, \sigma_2$ terms with real coefficients are allowed in the $cc$ and $ff$ blocks, and only real $\sigma_0$ and imaginary $\sigma_3$ terms are allowed in the $fc$ blocks. A further constraint arises from $C_3$ symmetry. Although any fixed strain breaks $C_3$, the full system remains symmetric if strain is also transformed
\bea
D[C_3] h(\mathcal{E},\mbf{k}) D[C_3]^\dag = h(C_3 \mathcal{E} C_3^{-1},C_3 \mbf{k})
\eea
 since the strain tensor transforms as $\mathcal{E}_l \to R^{-1}(2\pi/3) \mathcal{E}_l R(2\pi/3)$ as shown after \Eq{eq:strainsym}. This is equivalent to transforming $(\eps_{-}, \eps_{xy})^T \to C_3 (\eps_{-}, \eps_{xy})^T$ and $\eps_+ \to \eps_+$ in the Hamiltonian, placing constraints on the strain coupling in the heavy fermion basis. For example, we consider the $\Gamma_3$ block which transforms under the representation $D_{\Gamma_3}[C_3] = e^{\frac{2 \pi i}{3} \sigma_3}$ which is a vector representation. Since $\pmb{\sigma} = (\sigma_1, \sigma_2)$ also transforms in the vector representation of $D_{\Gamma_3}[C_3]$, the only symmetry-allowed combination of $\eps_{xy}, \eps_{-}$ is $c \, \pmb{\eps} \cdot \pmb{\sigma}$. Next we consider the $\eps_+$ term in this block, which is invariant under $C_3$ (it transforms as a scalar). Both $\sigma_0$ and $\sigma_3$ transform as a scalar, but both commute with $D_{\Gamma_3}[P] = - i \sigma_3$, which must anti-commute with all strain terms since strain preserves particle-hole. Thus only the vector representation is allowed at leading order in $\mbf{k}$. 
 
Repeating this procedure for the various blocks, we find the following parameterization of the perturbation $\delta H(\mathcal{E})$ to be 
\bea
\delta h(\mathcal{E}) = \bpm
c (\eps_{xy} \sigma_1 + \eps_- \sigma_2) & c' (\eps_{xy} \sigma_1 - \eps_- \sigma_2)  & i \gamma' \eps_+ \sigma_3 \\
c' (\eps_{xy} \sigma_1 - \eps_- \sigma_2) &  M' \eps_+  \sigma_2 & c'' (\eps_{xy} \sigma_0 - i \eps_- \sigma_3) \\
-i \gamma' \eps_+ \sigma_3& c'' (\eps_{xy} \sigma_0 + i \eps_- \sigma_3) & M_f (\eps_{xy} \sigma_1 + \eps_- \sigma_2) \\
\epm \ .
\eea
The coefficients may be found in the Main Text for the BM and KV model. Our approach has been to project the strain perturbation terms onto the original Wannier states and conduction electron wavefunctions at each $\mbf{k}$, which is degenerate first order perturbation theory. As a further approximation, we make a $k.p$ expansion and keep only $\mbf{k}$-independent terms. In the Main Text, we verify that the agreement between the strained BM model and the strained HF model is excellent, justifying this approximation.

We now terms which break the anti-commuting particle-hole symmetry $P$, which preserve all space group symmetries but break particle-hole. Such terms arise from non-local tunneling and relaxation effects in the KV model. In the HF model, the possible $P$-breaking (meaning they commute with $P$) and space-group preserving terms of low order in $\mbf{k}$ are found to be
\bea
\label{eq:Hrelaxedapp}
\delta h_{relaxed}(\mbf{k}) &= \bpm 
 \mu_1 \sigma_0+ v_2 (k_x \sigma_1 + k_y \sigma_2) & v_1 (k_x \sigma_1 - k_y \sigma_2)  &  0 \\
v_1 (k_x \sigma_1 - k_y \sigma_2) & \mu_2 \sigma_0 &  v_3(k_x  \sigma_0 - i k_y \sigma_3) \\
0 & v_3(k_x  \sigma_0 + i k_y  \sigma_3) & \mu_f \sigma_0 \\
\epm \\
\eea
consisting of the $c$-electron potentials $\mu_1, \mu_2$ and three velocities $v_1, v_2, v_3$. Without loss of generality, we have set the $f$-electron chemical potential to zero. The values of the coefficients may be found in the Main Text for the BM and KV models.

The parameters for strain and relaxation are calculated in the $\Gamma$ point approximation like the original HF parameters. Recall that the HF Hamiltonian $\delta h$ is computed from $\delta h_{\mu \nu}(\mbf{k}) = \braket{\mbf{k},\mu|H(\mbf{k})|\mbf{k},\nu}$ where $\mu,\nu$ index the $f$- and $c$-mode wavefunctions and $H(\mbf{k})$ is the BM or KV model in the plane wave basis. By construction, they are smooth and slowly varying around $\Gamma$. To compute any mass term (i.e. a $\mbf{k}$-independent term), we simply have $\delta h(\mbf{k}=0) =  \braket{\mbf{0},\mu|H(\mbf{0})|\mbf{0},\nu}$. Velocity terms are computed via $\pmb{\nabla}_k \delta h(\mbf{k})|_{\mbf{k}=0} =  \braket{\mbf{0},\mu|\pmb{\nabla}_k H(\mbf{0})|\mbf{0},\nu}$ which is an approximation since it neglects the wavefunction derivatives $\pmb{\nabla}_\mbf{k} \ket{\mbf{k},\nu}$. However, the smoothness of the wavefunctions ensures these derivatives are not large, and the approximation is confirmed numerically from the accuracy of the HF bands.

\section{Classical Theory of the Relaxation Field}
\label{app:relaxationcomputation}

In this Appendix, we discuss the classical elasticity theory from which the relaxation potential entering the KV model is obtained. We follow closely Refs. \cite{PhysRevB.107.075408} and  \cite{PhysRevB.107.075123}. 

The problem is set up as follows. We assume a long-wavelength continuum deformation field $\mbf{u}_{\ell}(\mbf{r})$ that describes the relaxation of the layers $\ell$. (There is no sublattice index since we are in the long wavelength limit, and we neglect corrugation such that $\del_z \mbf{u}_\ell = 0$.) The deformation field $\mbf{u}_\ell$ is solved for by minimizing the Hamiltonian
\bea
\label{eq:elasticdef}
H_0[\mbf{u}_\ell] &= \sum_\ell U_{elastic}[\mbf{u}_\ell] +  \sum_\ell  V[\mbf{u}_{\ell}-\mbf{u}_{\ell-1}] \\
&= \frac{1}{2} \int d^2r \, \sum_\ell \ \mathcal{K} (\pmb{\nabla} \cdot \mbf{u}_\ell)^2 + \mathcal{G} \big( (\del_x u_{x,\ell} - \del_y u_{y,\ell})^2 + (\del_x u_{y,\ell} + \del_y u_{x,\ell})^2 \big) \\
&\qquad + \int d^2r \, \sum_{\ell} V[\mbf{u}_{\ell} - \mbf{u}_{\ell-1}]
\eea
where the elastic (kinetic) energy term contains $\mathcal{K},\mathcal{G}$, the bulk/shear modulus of graphene and the inter-layer potential (adhesion) term is determined by the function
\bea
V[\mbf{u}] = \sum_\mbf{G} V_\mbf{G} \cos \mbf{G} \cdot \mbf{u}
\eea
which is an even, periodic function on the graphene lattice (here $\mbf{G}$ are the graphene reciprocal lattice vectors). A full list of energy functional parameters can be found in Ref. \cite{PhysRevB.107.075408} using the Wannier-based interpolation from DFT \cite{PhysRevB.93.235153}. 

We now assume a bilayer and define
\bea
\label{eq:hethom}
\mbf{u}(\mbf{r}) &= \mbf{u}_{top}(\mbf{r}) - \mbf{u}_{bottom}(\mbf{r}) \\
\mbf{u}_+(\mbf{r}) &= \frac{1}{2} \lp \mbf{u}_{top}(\mbf{r}) + \mbf{u}_{bottom}(\mbf{r}) \rp \\
\eea
which give the heterostrains and homostrains respectively. Since $U_{elastic}$ is diagonal and identical for all layers, the change of variables in \Eq{eq:hethom} is still decoupled (diagonal):
\bea
U_{elastic}[\mbf{u}_\ell] &= \frac{1}{2} U_{elastic}[\mbf{u}] + 2 U_{elastic}[\mbf{u}_+] \\
\eea
and hence $H$ is minimized by setting $U_{elastic}[\mbf{u}_+] = 0$. Thus we only need to consider the heterostrain Hamiltonian through
\bea
\label{eq:Hu}
H[\mbf{u}] &= \frac{1}{2} U_{elastic}[\mbf{u}] + V[\mbf{u}] \ .
\eea
The ansatz for $\mbf{u}$ takes the form of a rigid twist plus a moire-periodic relaxation (this ansatz can be generalized to include strain)
\bea
u_i(\mbf{r}) &=  \th \eps_{ij} r_j + \delta u_i(\mbf{r}) + O(\th^2), \qquad \delta u_i(\mbf{r}) = \delta u_i(\mbf{r}+\mbf{a}_M) \\
\eea
with summation implied and $\mbf{a}_M$ a moir\'e lattice vector. From this ansatz, we find that the rigid twist does not contribute kinetic energy in \Eq{eq:elasticdef}, so 
\bea
U_{elastic}[\mbf{u}] &= U_{elastic}[\delta\mbf{u}] 
\eea
which follows from the identities $\del_i \eps_{ij} r_j = 0$ (unsummed) and $\del_x \eps_{yj} r_j+\del_y \eps_{xj} r_j =0$. We rewrite the adhesion term via
\bea
V[u_i] = V[\th \eps_{ij} x_j + \delta u_i] &= \sum_\mbf{G} V_\mbf{G} \cos \lp \mbf{g}_\mbf{G}  \cdot \mbf{x} + \mbf{G} \cdot \delta \mbf{u} \rp, \qquad \th G_i \eps_{ij} x_j \equiv [g_\mbf{G}]_j x_j
\eea
using the moir\'e reciprocal vector $\mbf{g}$ mapped from the graphene reciprocal vector $\mbf{G}$ through $g_i = - \th \eps_{ij} G_j$ to leading order in $\th$. 

Having simplified the Hamiltonian, we minimize it using the Euler-Lagrange equations $\delta H = 0$. To derive the explicit form of this equation, it is helpful to rewrite the kinetic term $U_{elastic}$ in the form
\bea
\mathcal{K} (\pmb{\nabla} \cdot \mbf{u})^2 + \mathcal{G} \big( (\del_x u_{x} - \del_y u_{y})^2 + (\del_x u_{y} + \del_y u_{x})^2 \big) &= (\mathcal{K} +\mathcal{G}) \big( (\del_x u_x)^2  + (\del_y u_y)^2 \big) + 2 (\mathcal{K} - \mathcal{G}) \del_x u_x \del_y u_y \\
&\qquad + \mathcal{G} \big( (\del_x u_y)^2 + (\del_y u_x)^2  + 2 \del_x u_y \del_y u_x \big)  \\
&= (\mathcal{K} +\mathcal{G}) \big( (\del_x u_x)^2  + (\del_y u_y)^2 \big) + 2 \mathcal{K}  \del_x u_x \del_y u_y \\
&\qquad + \mathcal{G} \big( (\del_x u_y)^2 + (\del_y u_x)^2  \big) \ . \\
\eea
using $\del_x u_y \del_y u_x = \del_y u_y \del_x u_x$ up to total derivatives. Using another integration by parts in all terms, we get
\bea
\frac{1}{2} U_{elastic}[\mbf{u}] &= -\frac{1}{4} \int d^2r  \bpm u_x & u_y \epm \bpm (\mathcal{K} +\mathcal{G}) \del_x^2  + \mathcal{G} \del_y^2  & \mathcal{K} \del_x \del_y \\ \mathcal{K} \del_x \del_y & (\mathcal{K} +\mathcal{G}) \del_y^2  + \mathcal{G} \del_x^2 \\ \epm \bpm u_x \\ u_y \epm
\eea
The full Euler-Lagrange equation is then obtained from \Eq{eq:Hu} to be 
\bea
\label{eq:eom}
\delta H = 0 \implies - \frac{1}{2} \bpm (\mathcal{K} +\mathcal{G}) \del_x^2  + \mathcal{G} \del_y^2  & \mathcal{K} \del_x \del_y \\ \mathcal{K} \del_x \del_y & (\mathcal{K} +\mathcal{G}) \del_y^2  + \mathcal{G} \del_x^2 \\ \epm \bpm \delta u_x \\ \delta u_y \epm- \sum_\mbf{G} V_\mbf{G} \sin (\mbf{g}_\mbf{G} \cdot \mbf{r} + \mbf{G} \cdot \delta \mbf{u}) \bpm G_x \\ G_y \epm &= 0  \ .
\eea
To solve this equation, we use the fact that $\delta u(\mbf{r})$ is moir\'e-periodic and introduce the Fourier decomposition
\bea
\delta u(\mbf{r}) &= \sum_\mbf{g} \delta u(\mbf{g}) e^{i \mbf{g} \cdot \mbf{r}}, \qquad \sin (\mbf{g}_\mbf{G} \cdot \mbf{r} + \mbf{G} \cdot \delta \mbf{u}(\mbf{r})) = \sum_\mbf{g} f^{\delta u}_\mbf{G}(\mbf{g}) e^{i \mbf{g} \cdot \mbf{r}} \ .
\eea
In these variables, \Eq{eq:eom} can be written
\bea
 \frac{1}{2} \bpm (\mathcal{K} +\mathcal{G}) g_x^2  + \mathcal{G} g_y^2  & \mathcal{K} g_x g_y \\ \mathcal{K} g_x g_y & (\mathcal{K} +\mathcal{G}) g_y^2  + \mathcal{G} g_x^2 \\ \epm \bpm \delta u_x(\mbf{g}) \\ \delta u_y(\mbf{g}) \epm &= \sum_\mbf{G} V_\mbf{G} f^{\delta u}_\mbf{G}(\mbf{g}) \bpm G_x \\ G_y \epm \\
\eea
This equation can be solved iteratively starting from the initial guess $\delta \mbf{u} = 0$. The iterative scheme is
\bea
 \bpm \delta u^{(n+1)}_x(\mbf{g}) \\ \delta u^{(n+1)}_y(\mbf{g}) \epm &= 2 \bpm (\mathcal{K} +\mathcal{G}) g_x^2  + \mathcal{G} g_y^2  & \mathcal{K} g_x g_y \\ \mathcal{K} g_x g_y & (\mathcal{K} +\mathcal{G}) g_y^2  + \mathcal{G} g_x^2 \\ \epm^{-1} \sum_\mbf{G} V_\mbf{G} f^{\delta u^{(n)}}_\mbf{G}(\mbf{g}) \bpm G_x \\ G_y \epm 
\eea
where the graphene $\mbf{G}$ sum is truncated at 3 shells, and the moir\'e momentum $\mbf{g}$ is cutoff at 5 shells (or in general, however many are needed for the hoppings in the electron Hamiltonian). The resulting strain profile can be plugged into the gradient expansion developed in Refs. \cite{PhysRevB.107.075408} and  \cite{PhysRevB.107.075123}, which we refer the reader to, in order to reproduce the atomistic electronic structure within a low-energy continuum model.

\end{document}